\documentclass[letter, 10pt, conference]{ieeeconf}      

\IEEEoverridecommandlockouts                              
\usepackage{graphics} 
\usepackage{epsfig} 
\usepackage{amsmath} 
\usepackage{amssymb}  
\usepackage{bm}
\usepackage{color}
\definecolor{dark-gray}{gray}{0.45}
\usepackage{tikz}
\usepackage{url}
\usepackage{algpseudocode}
\usepackage{algorithm}
\usepackage{subfigure}
\usepackage{caption}
\usepackage{dsfont} 
\usepackage{booktabs,array}
\usepackage{graphicx}
\usepackage{mathtools}
\usepackage{multicol}
\usepackage{multirow}
\usepackage{mathtools,bbm}
\usepackage{marvosym}
\usepackage{blindtext}
\usepackage{comment}
\usepackage{siunitx}
\usepackage{verbatim}
\usepackage{hyperref}
\usetikzlibrary{shapes,snakes,3d,calc,arrows}

\usepackage[normalem]{ulem}

\newcommand{\hcite}[1]{\hyperref[cit_bib]{\cite{#1}}}

\DeclarePairedDelimiter{\ceil}{\lceil}{\rceil}

\newtheorem{Prop}{\textbf{Proposition}}

\include{tikzDefs}

\title{\LARGE \bf
{Distributed Strategies for Dynamic Coverage\\ with Limited Sensing Capabilities}
}

\author{Marco Fabris and Angelo Cenedese
\thanks{\noindent A. Cenedese, and M. Fabris {\tt\small \{angelo.cenedese@unipd.it, marco.fabris.7@studenti.unipd.it\}} are with the Department of Information Engineering, University of Padova, Italy.}
}

\begin{document}

\maketitle
\thispagestyle{empty}
\pagestyle{empty}

\begin{abstract}
	In this work, it is presented the development of a novel distributed algorithm performing robotic coverage, clustering and dispatch around an event in static-obstacle-structured environments without relying on metric information. Specifically, the aim is to account for the trade-off between local communication given by bearing visibility sensors installed on each agent involved, optimal deployment in closed unknown scenarios and focus of a group of agents on one point of interest. The particular targets of this study can be summarized as 1. the computation, under certain topological assumptions, of a lower bound for the number of required agents, which are provided by a realistic geometric model (e.g. a round shape) to emphasize physical limitations;
	2. the minimization of the number of nodes and links maintaining a distributed approach over a connected communication graph; 3. the identification of an activation cluster around an event with a radial decreasing intensity, sensed by each agent; 4. the attempt to send the agents belonging to the cluster towards the most intense point in the scenario by minimizing a weighted isoperimetric functional. 
\end{abstract}

\section{Introduction}
In the recent decades, researchers have focused their attention in multi-agent networks \hcite{MesbahiEgerstedt2010} composed by a large number of smart devices that can collaborate to achieve a common goal and are capable of taking local decisions independently, without any supervisory framework. Although complex large-scale monitoring and control systems are not new, as, for example, urban traffic control \hcite{WangDjahelMcManis2014} or smart grids applications \hcite{GaoXiaoLiuLiangChen2012}, novel architectural models are still emerging, as 
UAVs for multimedia delivery in industry \cite{AlTurjman2018}. The choice of multi-agent systems is motivated by the fact that they can be supported by hierarchical developments, 
which has the considerable advantage to be relatively easy to be designed maintaining safety guarantees. However, these systems require reliable sensors and actuators that are, generally, very expensive. Furthermore, the potential adoption of a centralized approach has been widely proven to be a limitation for scalability, flexibility, adaptivity and maintenance \hcite{GeYangHan2017}. The recent trend, indeed, encourages the substitution of costly sensors, actuators and communication mechanisms with a larger number of devices that can autonomously compensate potential failures and a high computational burden through smart communication protocols and strong cooperation. 

\subsubsection*{Related works}
Classic coverage applications are based on centralized localization for each robot (e.g. GPS) in order to accomplish equitable partitions of the environment (e.g. Voronoi tessellations) by minimizing a coverage cost \hcite{DuFaberGunzburger1999}. However, such approaches need a priori information about the environment where robots are deployed. Lately, complete sensor coverage of indoor environments using swarms of robots~has~been~studied~\hcite{RutishauserCorrellMartinoli2009}-\hcite{DudekJenkinMilosWilkes1991}~by~modeling~the~known~network as a graph and assuming robots to have global localization, with the possibility to navigate independently from one location to another in a global coordinate frame. In all of these lines of research, global localization of the robots, a priori knowledge of the environment (obstacle configuration), availability of metric information and ability to control the robots from one point in the environment to another have been presumed.
In recent years, coverage by sensor network is studied more formally by using Simplicial Complex Theory \hcite{Zomorodian2010} and homological tools from algebraic topology, as in \hcite{SilvaGhrist2006}-\hcite{GhristLipskyDerenickSperanzon2012}-\hcite{SilvaGhrist2007}, requiring little to no metric information. 
One of the latest research, from which our work draws inspiration, is \hcite{RamaithitimaWhitzerBhattacharyaKumar2015}. Here, authors offer new perspectives improving the previous studies by using a different approach based on three aspects. Firstly, their method is robust to agent failure at any configuration. Secondly, the approach of pushing agents through the network links in order to expand the graph frontier, instead of navigating,  does not require any restriction on the minimum distance between two neighbors. Lastly, their algorithm does not demand the workspace to be simply-connected, as demonstrated via experimental results.

\subsubsection*{Contribution and outline of the paper}
Starting from \hcite{RamaithitimaWhitzerBhattacharyaKumar2015}, we implement a more advanced and fully decentralized algorithm to fulfill coverage and focus on event, motivated by several works in the fields of event detection \hcite{WitteburgDziengelAdlerKasmi2012}, cluster selection \hcite{MehdiTayarani2014} and robotic dispatch \hcite{LukicStojmenovic2013}. A completely distributed approach, i.e. requiring local communication only, is presented: each agent needs to share with its neighbors just the information about measured bearing angles and neighborhoods. In particular, useless agents are sought by exploiting bearing angles, instead of computing graph cycles. Moreover, geometric models for the agents are adopted, e.g. circular shapes: this leads us to take into account collisions and space occupation in order to provide a lower bound for the number of agents deployed by our algorithm on 
virtual scenarios with rectangular features. Once coverage is completed, the focus on a preset event is then attained: in this final procedure each agent builds on a network deformation that needs to increase the event information while not losing connectivity and enhancing communication among activated companions. The remainder of this paper is organized as follows. In Sec. \ref{sec:mathematicalPreliminaries}, some preliminary definitions and mathematical models are introduced, while Sec. \ref{sec:algorithm_design} illustrates central ideas for the implementation of our algorithm. The discussion continues in Sec. \ref{sec:number_deployed_agents}
 with the analysis of lower and upper bounds for the minimum number of deployed agents.
In addition, with the numerical results reporting performances and limitations of our algorithm that can be found in Sec. \ref{sec:numerical_simulations}. Finally, future directions and conclusions are drawn in Sec. \ref{sec:conclusions}.

\section{Preliminaries} \label{sec:mathematicalPreliminaries}

\subsection{Basic notations and definitions}
In every part of this paper, letters $i,j,k$ and their variants are used as indexes; while letter $l$ and its variants indicate finite lengths of vectors or number of elements in a sequence.\\
Let $\Omega$ be a set. We indicate with $\mathrm{cl}(\Omega)$ its closure and $\mathrm{int}(\Omega)$ its interior. Whenever $\Omega$ is discrete and finite or empty we denote its cardinality with $\left|\Omega \right| \in \mathbb{N}$, otherwise $\left|\Omega \right| = +\infty$.\\
Let $\mathbf{p}_{k} = (x_{k},y_{k}) \in \mathbb{R}^{2}$ be a point. We define
\begin{itemize}
	\item a segment $st_{ij}$ with vertices $(\mathbf{p}_{i},\mathbf{p}_{j})$ as the collection of points $ \left\lbrace \omega \mathbf{p}_{i} +(1-\omega) \mathbf{p}_{j} ~|~\mathbf{p}_{i} \neq \mathbf{p}_{j},~\forall \omega \in [0,1] \right\rbrace$;
	\item a polygonal line $pl_{[0,l]}$ as the collection of segments $ \left\lbrace st_{k_{0}k_{1}}, st_{k_{1}k_{2}},..., st_{k_{l-1}k_{l}} \right\rbrace =: \left\lbrace st_{1}, st_{2},..., st_{l} \right\rbrace$, assuming $l\geq 2$ finite, and we say that a polygonal line is closed whenever $\mathbf{p}_{k_{0}} = \mathbf{p}_{k_{l}}$, using the notation $pl_{[1,l]}$ in the latter case;
	\item a polygon figure $pn_{[1,l]}$ as the union $ \bigcup_{k=1}^{l} st_{k}$, such that each segment $st_{k}$ belongs to a closed polygonal line $pl_{[1,l]} $ that satisfies the following two properties:
	\begin{itemize}
		\item $\mathbf{p}_{k_{i}} \neq \mathbf{p}_{k_{j}}$ for all $i,j = 1,...,l$, with $i\neq j$;
		\item $\sum_{i< j;~ i,j = 1,...,l} \left|  st_{i}  \cap  st_{j} \right| = l $. 
	\end{itemize} 
\end{itemize}
Whenever $\Omega_{k}$ represents either a segment or a polygon, we denote with $\Omega_{k}^{2} \subset \mathbb{R}^{2}$ the surface of $\Omega_{k}$, setting $\Omega_{k}^{2} = \Omega_{k}$ if and only if $\Omega_{k}$ is a segment, such that if $\Omega = \bigcup_{k=1}^{l} \Omega_{k}$ then $\Omega^{2} = \bigcup_{k=1}^{l} \Omega^{2}_{k}$. We indicate with $\left\|\Omega_{k} \right\|_{1} > 0$ and $\left\|\Omega_{k} \right\|_{2} \geq 0$ the perimeter of $\Omega_{k}$ and the area of $\Omega_{k}^{2}$ respectively, setting $\left\|\Omega_{k} \right\|_{2} = 0$ if and only if $\Omega_{k}$ is a segment. Furthermore, we assume that, if $\mathrm{int}(\Omega_{i}) \cap \mathrm{int}(\Omega_{j}) = \emptyset$ for all $i,j=1,...,l$, with $i\neq j$, then $\left\|\Omega \right\|_{\bar{k}} = \sum_{k=1}^{l} \left\|\Omega_{k}\right\|_{\bar{k}}$ for $\bar{k}=1,2$.

\subsection{Mathematical models}
In this subsection, we present the main assumptions for all the models adopted in the network.

\subsubsection*{Scenario}

We assume that a scenario $SC = (EN,OB)$ is the planar space where actions take place, composed of an enclosure $EN$ and, possibly, a set of obstacles $OB=\bigcup^{l_{ob}}_{k=1} ob_{k}$ with finite number $l_{ob}$ of physical barriers $ob_{k}$. Whenever a scenario is obstacle-free we just assume that $SC = EN$. For the sake of simplicity, we impose that $EN$ is a polygon and each obstacle $ob_{k}$ is either a polygon or a segment. Moreover, to characterize $SC$ as a realistic environment, we also impose the following properties on its potential obstacles:
\begin{itemize}
	\item $\mathrm{int}(ob_{i}) \cap \mathrm{int}(ob_{j}) = \emptyset$ for all $i,j=1,...,l_{ob}$, $i\neq j$;
	\item $\mathrm{cl}(OB^{2}) \subset \mathrm{int}(EN^{2})$;
	\item $ \left\|OB\right\|_{2} \leq k_{SC} \left\|EN\right\|_{2}$ for a given $k_{SC}\in (0,1)$.
\end{itemize}
These three properties state that obstacles cannot overlap each other, they must be contained inside the enclosure, and their space occupation in the enclosure must be reasonably low with respect to the total available space. Finally, we define the surface to be covered as $CS^{2} = \mathrm{int}( EN^{2} ) \setminus \mathrm{cl}(OB^{2})$ and the scenario boundaries as $SB = EN \cup OB$.
\subsubsection*{Event}
An event is a point $EV \in EN^{2}$ that becomes significant inside the scenario after a complete coverage has been already attained and with a relevance that is radially decreasing. The latter is modeled by the real scalar function
\begin{equation}
f_{EV}(\mathbf{p}) = k_{EV} \exp\left(-\left\|\mathbf{p}-EV\right\|^{2}/r^{2}_{EV}\right), \quad \forall \mathbf{p}\in EN^{2}
\end{equation}
where $k_{EV} > 0$ and $r_{EV}>0$ represent the maximum intensity and the decay distance respectively.
\subsubsection*{Multi-agent system}
According to the existing literature, a $n$-agent system can be modeled through a graph $\mathcal{G}=\left(\mathcal{V},\:\mathcal{E}\right)$ so that each element in the nodes set $\mathcal{V}=\left\{v_1 \dots v_n\right\}$ corresponds to an agent in the group, while the edge set $\mathcal{E}\subseteq \mathcal{V}\times \mathcal{V}$ describes the agents interactions. 
In the rest of the paper, we assume that $\mathcal{G}$ is undirected and that the set $\mathcal{E}$ depicts both agents visibility and communication capabilities, meaning that there exists $e_{ij}=(v_i , v_j) \in \mathcal{E}$ if and only if the $i$-th and $j$-th
agents can sense each other and are able to
reciprocally exchange information according to some predetermined communication protocol.
The agents interplays are generally represented by the \textit{adjacency matrix} $\mathbf{A} \in \mathbb{R}^{n \times n}$ such that $[\mathbf{A}]_{ij}=1$ if $e_{ij}\in\mathcal{E}$ (i.e., $v_i$ and $v_j$ are adjacent) and $[\mathbf{A}]_{ij}=0$ otherwise.
For each node $v_i$ in $\mathcal{G}$, the set $\mathcal{N}_i=\left\{v_j\in\mathcal{V}\;|\; [\mathbf{A}]_{ij}=1\right\} \subseteq \mathcal{V}$, named \textit{neighborhood}, thus represents the set of agents interacting with the $i$-th agent. By convention, it holds that $v_i \notin \mathcal{N}_i$ and, whenever $\mathcal{N}_i$ possesses any order, $v_j = \left[\mathcal{N}_i\right]_{k}$ is the $k$-th neighboring vertex of $v_i$, for $k=1,...,|\mathcal{N}_i|$. The cardinality of $\mathcal{N}_i$ is the \textit{degree}, $\text{deg}(v_i)$, of the $i$-th agent. 
Vertices and edges can be both weighted by positive real weights, denoted with $|v_{i}|$ and $|e_{ij}|$ respectively (if nothing it is said, then $|v_{i}|=1$ and $|e_{ij}|=1$ is the default choice). Thus, the degree of a vertex $v_i$ can be redefined as the edge degree $\text{deg}_{e}(v_i) = \sum_{j=1}^{\text{deg}(v_{i})} |e_{ij}|$ or the vertex degree $\text{deg}_{v}(v_i) = |v_i|\text{deg}(v_i)$. Generally, given the generic positive weights $\text{w}_{1},...,\text{w}_{n}$, we define the weighted volume of the graph $\mathcal{G}$ as $\text{vol}_{\text{w}}(\mathcal{G})=\sum_{i=1}^{n}\text{deg}_{\text{w}}(v_i)$. Finally, for any subgraph $\mathcal{G}_{S}:=(\mathcal{V}_{S},\mathcal{E}_{S})\subseteq (\mathcal{V},\mathcal{E})$, we define its complement $\mathcal{G}_{\overline{S}} = (\mathcal{V}_{\overline{S}},\mathcal{E}_{\overline{S}}) := \mathcal{G}\setminus \mathcal{G}_{S}$ and the cut between $\mathcal{G}_{S}$ and $\mathcal{G}_{\overline{S}}$ as the set $\partial\mathcal{E}_{S} = \partial\mathcal{E}_{\overline{S}} := \left\lbrace (v_i,v_j) ~|~ v_i \in \mathcal{V}_{S} ~\&~  v_j \in\mathcal{V}_{\overline{S}} \right\rbrace \subseteq \mathcal{E}$.
\subsubsection*{Agents}
Any agent $a_{k}=(v_{k},\mathbf{p}_{k},r_{b},r_{v})$ is represented by a vertex in the graph $\mathcal{G}$ and modeled by a circle with radius\footnote{Because of this fact, corridors in the scenario with a width lesser than the body diameter $2r_{b}$, can never be accessed by agents.} $r_{b}>0$ centered in $\mathbf{p}_{k}\in CS^{2}$. They are also provided with a camera installed on the center with a visibility radius $r_{v}>0$ in order to acquire bearing measurements (\hcite{RamaithitimaWhitzerBhattacharyaKumar2015}, Sec. II-A). These are used to measure the angles that agents create with one or more neighbors. A bearing measurement (in radians) of an agent $a_{k}$ w.r.t. one of its neighbor $a_{i}$ is denoted with $\theta^{k}_{i}\in [-\pi,\pi)$. If it measures the bearing to another agent $a_{j}$ as $\theta^{i}_{j}$, then we define the bearing to $a_{k}$ relative to the bearing to $a_{j}$ as $\theta^{k}_{ij} = ((\theta^{k}_{i}-\theta^{k}_{j}) \mod 2\pi) - \pi$, such that $\theta^{k}_{ij}\in [-\pi,\pi)$. Two agents $(a_{i},a_{j})$ are said neighbors if:
\begin{itemize}
	\item $\left\| \mathbf{p}_{i}-\mathbf{p}_{j}\right\| \leq r_{v}$, i.e., their centers are not distant more than the visibility radius $r_{v}$;
	\item $(st_{ij} \cap CS^{2}) \subset CS^{2}$, i.e., there is no barrier in between;
\end{itemize}
We assume that $a_{i}$ and $a_{j}$ communicate even though between them another agent has already been placed. Furthermore, agents are provided with contact sensors that are triggered when touching any barrier in the scenario or another agent, able to roughly estimate the direction of an impact (\hcite{RamaithitimaWhitzerBhattacharyaKumar2015}, Sec. II-C) within an error $\Delta \tau = \pi / N_{T}$, where $N_{T}$ is the number of contact points in each touch sensor. Finally, each agent is provided by an event sensor in order to sense an estimate $\hat{f}_{EV}(\mathbf{p})$ of the event intensity. \\
Each agent is controlled using the bearing-based visual homing controller (\hcite{RamaithitimaWhitzerBhattacharyaKumar2015}, Sec. II-D), meaning that the velocity $\dot{\mathbf{p}}_{i}$ of robot $a_{i}$ is given by $\dot{\mathbf{p}}_{i} = k_{p}\sum_{v_j\in\mathcal{M}_{i}} (\theta^{i}_{j,des}-\theta_{j}^{i})$, where $\mathcal{M}_{i} \subset \mathcal{N}_{i}$ is the list of robots that are neighbors of $a_{i}$ and which can be used as landmarks, $\theta^{i}_{j,des}$ is the desired bearing with landmark $a_{j}$ and $k_{p}$ is a feedback gain.
\subsubsection*{Base station} 
A base station is a point $BS=\mathbf{p}_{1} \in CS^{2}$ that generates all robots during coverage and it also represents the position of the first agent $a_{1}$. Because of this choice, we assume that $a_{1}$ cannot be removed while seeking potential redundant agents in the network (see Subsec. \ref{subsec:coverage_policy}).

\vspace{0.1cm}
\section{Algorithm design} \label{sec:algorithm_design}
A general overview of the main procedure is illustrated in Alg. \ref{alg:outline} and described as follows.
\begin{itemize}
	\item \textit{Coverage}: Firstly, agents are deployed until full coverage is attained (line \ref{line:coverage}).
	\item \textit{Clustering}: Subsequently, they sense an event $EV$ in the scenario and a cluster $\mathcal{G}_{CL}\subseteq \mathcal{G}$ with preselected cardinality $n_{CL}=\left|\mathcal{G}_{CL}\right|$ is created (line \ref{line:clustering}). The cluster formation begins from a leader vertex $v^{\star}$ that measures the highest intensity: this node can be elected by means of max-consensus algorithms \hcite{OlivaSetola2013} (line \ref{line:max-consensus}). Then, the cluster grows with a greedy approach, maintaining its connectivity.
	\item \textit{Dispatch}:  Finally, agents belonging to $\mathcal{G}_{CL}$ perform a dispatch (line \ref{line:dispatch}) according to the minimization of an isoperimetric functional $h_{\mathcal{G}}(\mathcal{G}_{CL})$ with the purpose to drive the cluster's elements close to the event origin $EV$ as far as possible, maintaining the connections constituted right after the coverage. This minimization takes place iterating on cluster $\mathcal{G}_{CL}$, where a maximum number of iterations\footnote{Iterations are counted by variables $c_{di}$, stored in each node $v_i\in \mathcal{V}_{CL}$.} $MaxIter$ is fixed and each session can disable the flag $\text{f}_{d}^{\star}$, breaking the dispatch loop. We also decide to adaptively choose the leader node $v^{\star} $ (line \ref{line:max_con}), since, at each cycle, the event sensing for cluster nodes may vary while they move.
\end{itemize}

\subsection{Coverage stage}\label{subsec:coverage_policy}
As proposed in \hcite{RamaithitimaWhitzerBhattacharyaKumar2015}, let us denote the Vietoris-Rips Complex of the set of the deployed agents\footnote{Abuse of notation: a simplicial complex depends on the algorithm status.} with $\mathcal{R}_{r_{v}}$, the frontier subcomplex with $\mathcal{F}\subseteq\mathcal{R}_{r_{v}}$, the obstacle subcomplex with $\mathcal{O}\subseteq\mathcal{R}_{r_{v}}$ and the fence subcomplex with $\mathcal{K}=\mathcal{F}\cup \mathcal{O}$. To accomplish this stage, we implement the \textsc{Coverage} function at line \ref{line:coverage} in Alg.\ref{alg:outline} adopting their hexagonal-packing-based coverage algorithm. However, since the relative homology $H_{2}(\mathcal{R}_{r_{v}},\mathcal{K})$ is required to be stored in a centralized \textcolor{white}{manner -------------------------------------------------------------------------------}
\vspace{-1.1cm}
\begin{algorithm}
	\caption{Outline of the main procedure}\label{alg:outline}
	\begin{algorithmic}[1]
		\small
		\State $\mathcal{G}\leftarrow$\textsc{Coverage}(); \label{line:coverage}
		\For{each agent $a_i$, s.t. $i=1,...,n$}
		\State $|v_{i}| \leftarrow \hat{f}_{EV}(\mathbf{p}_{i})$;
		\EndFor
		\For{all $e_{ij}\in \mathcal{E}$}
		\State $|e_{ij}| \leftarrow (|v_{i}|+|v_{j}|)/2$; \label{line:edge_weighting}
		\EndFor
		\State $v^{\star}\leftarrow$\textsc{Max-Consensus}$(\mathcal{G},$BS$)$; \label{line:max-consensus}
		\State $\mathcal{G}_{CL}\leftarrow \left\lbrace v^{\star}\right\rbrace$
		\State \textsc{Clustering}($v^{\star}$,$1$); \label{line:clustering}
		\For{all nodes $v_i \in \mathcal{G}_{CL}$}
		\State $[c_{di},\text{f}_{di}] \leftarrow [0,\mathbf{false}]$;
		\EndFor
		\While{$ c_{d}^{\star} < MaxIter $ \textbf{and} $\text{f}_{d}^{\star} = \mathbf{false}$}
		\State $v^{\star}\leftarrow$\textsc{Max-Consensus}$(\mathcal{G}_{CL},v^{\star})$; \label{line:max_con}
		\State $\begin{bmatrix}
		c_{d}^{\star}, \text{f}_{d}^{\star}
		\end{bmatrix}\leftarrow$\textsc{Dispatch}($v^{\star}$,$c_{d}^{\star}+1$,$\mathbf{true}$); \label{line:dispatch}
		\EndWhile
	\end{algorithmic}
\end{algorithm} 

\noindent manner (\hcite{RamaithitimaWhitzerBhattacharyaKumar2015}, Sec. II-B), that procedure is not completely distributed. To account for this aspect, we introduce an improvement to this scheme: conversely to the homology-based approach utilized in \hcite{RamaithitimaWhitzerBhattacharyaKumar2015}, Sec. II-E, we pinpoint a redundant agent in the network exploiting reciprocal bearing measurement values. This allows us to rule out agents belonging to the interior of a 1-simplex (a segment)  or 2-simplex (a triangle) taking advantage of simple angle properties\footnote{The remarkable facts that the summation result of the three convex angles in any triangle is equivalent to a straight angle allows us to identify 1-simplex-redundant agents. Moreover, 2-simplex-redundant agents are spotted whenever three explementary angles with the same vertex exist.}. 
In Alg. \ref{alg:useless_agents}, we examine these two different cases considering each triplet of distinct agents $\boldsymbol{a}_{k}=(a_{k_{1}},a_{k_{2}},a_{k_{3}})$ that forms a 2-simplex and determine whether an agent is redundant to coverage purposes.

\begin{algorithm}
	\caption{Redundant agent search}\label{alg:useless_agents}
	\begin{algorithmic}[!]
		\small
		\For{$k=1,2,...$ s.t. $\mathbf{a}_{k}$ is a 2-simplex} 
		\State  $\boldsymbol{\theta}_{k} \leftarrow \begin{bmatrix}
		\theta_{k_{2}k_{3}}^{k_{1}} & \theta_{k_{1}k_{3}}^{k_{2}} & \theta_{k_{1}k_{2}}^{k_{3}}
		\end{bmatrix}$;
		\If{$\exists i \in \left\lbrace1,2,3 \right\rbrace$ s.t. $\left|[\boldsymbol{\theta}_{k}]_{i}\right| = \pi$ \textbf{and} $k_{i}\neq 1$}
		\State label $a_{k_{i}}$ as 1-simplex-redundant;
		\EndIf
		\State $\mathcal{N}_{k\cap} \leftarrow \mathcal{N}_{k_{1}} \cap \mathcal{N}_{k_{2}} \cap \mathcal{N}_{k_{3}}$;
		\For{$j=1,...,\left|\mathcal{N}_{k\cap}\right|$ s.t. $a_{k_{j}} \in \mathcal{N}_{k\cap}$}
		\If{$\left| \theta_{k_{1}k_{2}}^{k_{j}} +\theta_{k_{2}k_{3}}^{k_{j}}+\theta_{k_{3}k_{1}}^{k_{j}}\right|= 2\pi$ \textbf{and} $k_{j}\neq 1$}
		\State label $a_{k_{j}}$ as 2-simplex-redundant;
		\EndIf
		\EndFor
		\EndFor
	\end{algorithmic}
\end{algorithm}
$~$\vspace{-0.5cm}  

\subsection{Clustering stage}
In this intermediate stage, illustrated in Alg. \ref{alg:clustering}, we suppose that each node in the network has already sensed the event intensity, and hence each edge weight in the graph $\mathcal{G}$ has already been assigned by a specific function of the measurements (e.g. the function\footnote{This weighting function has to be chosen according to condition written on the r.h.s. at line \ref{line:new_leader} in Alg. \ref{alg:clustering}} at line \ref{line:edge_weighting} in Alg. \ref{alg:outline}): it follows that each neighborhood $\mathcal{N}_{i}$ in $\mathcal{G}$ can be sorted in a descending order according to each weight $|v_j|$, where $v_j = \left[\mathcal{N}_{i}\right]_{k}$, $k=1,...,|\mathcal{N}_{i}|$. The latter assumption contributes to implement a greedy approach: each agent $v_i$ is labeled as a cluster node if it is chosen as the best option in the neighborhood $\mathcal{N}_i$ according to criteria that reflect local optimality. Alg. \ref{alg:clustering} should be actually seen as a distributed one-hop loop and not as a recursive procedure, since it is not required to save the whole state of an agent in its stack while hopping. Keeping track of the last frame by including three counters and two additional flags -- used to report the affiliations to cluster $\mathcal{G}_{CL}$ and the subgraph of completed vertices $\mathcal{G}_{CO} \subseteq \mathcal{G}$ -- is enough to preserve information. Indeed, for every agent $a_{i}$, counters $\bar{k}_{i}$, $k_i$, $|\mathcal{G}_{CL}|$ are just counting increasing quantities and the two aforementioned flags, implementing lines \ref{line:cl_grows} and \ref{line:co_grows}, would be set only once. Moreover, when the \textsc{Clustering} function is invoked to perform a communication hop from $v_i$ to $v_j$ in a real framework, only the information related to the cardinality $|\mathcal{G}_{CL}|$ has to be passed and updated.\\
Three steps can be identified in Alg. \ref{alg:clustering}, which generally stops once $|\mathcal{G}_{CL}|=n_{CL}$ holds (line \ref{line:stop}). Firstly, for $\bar{k}_i = 1$, all vertices in neighborhood $\mathcal{N}_i$ are inserted into cluster $\mathcal{G}_{CL}$ (line \ref{line:cl_grows}), as far as possible: if a new leader $v_j$ is found (line \ref{line:new_leader}) a hop is performed according to a greedy paradigm. Secondly, for $\bar{k}_i = 2$, nodes in $\mathcal{N}_i$ are forcefully visited in order to expand the cluster, in case condition $|\mathcal{G}_{CL}|=n_{CL}$ is yet to be attained. If the latter is not verified at the end of this loop then vertex $v_i$ is labeled as complete (line \ref{line:co_grows}); namely, all nodes in $\left\lbrace v_i \cup \mathcal{N}_i \right\rbrace$ already belong to cluster $\mathcal{G}_{CL}$. Lastly, for $\bar{k}_i = 3$, vertex $v_i$ is definitively left and its non-complete neighboring nodes are selected (line \ref{line:nnn}) to be visited again.
$~$\vspace{-0.5cm}
\begin{algorithm}
	\caption{Focus on event: cluster formation\\ \textbf{procedure} \textsc{Clustering}$\left(v_{i},\left|\mathcal{G}_{CL}\right|\right)$}\label{alg:clustering}
	\begin{algorithmic}[1]
		\small
		\For{$\bar{k}_{i}=1,2,3$}
			\State $k_i \leftarrow 0$;
			\While{$k_i < \deg(v_{i})$ \textbf{and} $\left|\mathcal{G}_{CL}\right| < n_{CL}$}
				\State $k_i\leftarrow k_i+1$;
				\State $v_{j} \leftarrow \left[\mathcal{N}_{i}\right]_{k_{i}}$;
				\If{$v_{j} \notin \mathcal{G}_{CL} $ \textbf{or} $\bar{k}_i=3$}
					\If{$\bar{k}_{i} < 3$}
						\State $\mathcal{G}_{CL} \leftarrow\mathcal{G}_{CL} \cup \left\lbrace v_{j}\right\rbrace$; \label{line:cl_grows}
					\EndIf
					\State $k_{i1} \leftarrow $ $\bar{k}_{i}=1$ \textbf{and} $|v_j| > |e_{ij}|$; \label{line:new_leader}
					\State $k_{i2} \leftarrow $ $\bar{k}_{i}=2$;
					\State $k_{i3} \leftarrow $ $\bar{k}_{i}=3$ \textbf{and} $v_j \notin \mathcal{G}_{CO}$; \label{line:nnn}
					\If{$\left|\mathcal{G}_{CL}\right| < n_{CL}$ \textbf{and} ($k_{i1}$ \textbf{or} $k_{i2}$ \textbf{or} $k_{i3}$)} \label{line:stop}
						\State \textsc{Clustering}$\left(v_{i},\left|\mathcal{G}_{CL}\right|\right)$; 
					\EndIf
				\EndIf
				\If{$\bar{k}_{i}=2$ \textbf{and} $k_i=\deg(v_i)$}
					\State $\mathcal{G}_{CO} \leftarrow\mathcal{G}_{CO} \cup \left\lbrace v_i\right\rbrace$; \label{line:co_grows}
				\EndIf
			\EndWhile
		\EndFor
	\end{algorithmic}
\end{algorithm} 
$~$\vspace{-0.5cm}

\subsection{Dispatch stage}
In this final stage, agents belonging to cluster $\mathcal{G}_{CL}$ are dispatched as shown in Alg. \ref{alg:dispatch}. Given a temporary leader $v_i$ and its restricted neighborhood $\mathcal{N}_i^{\star}$ (defined at line \ref{line:restricted_neigh}) sorted decreasingly w.r.t. the event intensity, each node $v_j \in \mathcal{N}_i^{\star}$ is sent\footnote{We assume the bearing-based visual homing controller follows a suitable feedback control law to perform this navigation, e.g. steering the bearing measurement $\theta^{i}_{j}$ to zero while $a_j$ is moving.} towards $v_i$ itself from position $\mathbf{p}_j(t)$ to $\mathbf{p}_j(t+T) = \mathbf{p}_j(t) + K_{c_{di},j}(\mathbf{p}_i(t)-\mathbf{p}_j(t))/\left\| \mathbf{p}_i(t)-\mathbf{p}_j(t)\right\|$, $K_{c_{di},j}\geq 0$, as far as collisions or visibility issues do not arise (line \ref{line:collision_visib_issues}), i.e. $K_{c_{di},j}=0$. Moreover, a further stopping criterion (line \ref{line: stop_cond_deltavol}) for agent $a_j$ is given by the impossibility to locally minimize
\begin{equation}\label{eq:isoperimetric_funct}
h_{\mathcal{G}}(\mathcal{G}_{CL}) := \underset{h_{CL}}{\underbrace{\dfrac{|\partial\mathcal{G}_{CL}|}{\mathrm{vol}_v(\mathcal{G}_{CL})}} }+ \underset{h_{\overline{CL}}}{\underbrace{\dfrac{|\partial\mathcal{G}_{\overline{CL}}|}{\mathrm{vol}_v(\mathcal{G}_{\overline{CL}})}}}.
\end{equation}
Drawing inspiration from isoperimetric problems \hcite{Chung1996}, functional in \eqref{eq:isoperimetric_funct} is chosen to represent the \textquotedblleft bottleneckedness\textquotedblright$~$of the graph $\mathcal{G}_{CL}$ w.r.t. subgraph $\mathcal{G}$: the higher $h_{\mathcal{G}}(\mathcal{G}_{CL})$ the interconnected $\mathcal{G}_{CL}$ appears inside $\mathcal{G}$. To minimize $h_{\mathcal{G}}(\mathcal{G}_{CL})$, we choose not to add new edges from nodes in $\mathcal{G}_{CL}$ to nodes in $\mathcal{G}_{\overline{CL}}$: keeping $\partial\mathcal{E}_{CL}= \partial\mathcal{E}_{\overline{CL}}$ constant during the dispatch procedure allows us to consider term $h_{CL}$ only, since $\mathrm{vol}_v(\mathcal{G}_{\overline{CL}})$ cannot vary and, thus, to simplify the minimization. Defining the positive quantities $\epsilon_{S}:=\mathrm{vol}_{v}(\mathcal{G}_{CL})$, $\epsilon_{\overline{S}}:=\mathrm{vol}_{v}(\mathcal{G}_{\overline{CL}})$ and $\epsilon_{C}:=|\partial\mathcal{G}_{CL}|$, this heuristics is justified by the fact that the isoperimetric functional variation
\begin{align}
\Delta h_{\mathcal{G}}(\mathcal{G}_{CL}) = -\epsilon_{C}(\epsilon_{S}^{-1}+\epsilon_{\overline{S}}^{-1}) + \nonumber\\  (\epsilon_{C}+\Delta\epsilon_{C})\left[(\epsilon_{S}+\Delta\epsilon_{S})^{-1}+(\epsilon_{\overline{S}}+\Delta\epsilon_{\overline{S}})^{-1}\right]  
\end{align}
is expected not to become positive in all frameworks of interest. Condition $\Delta h_{\mathcal{G}}(\mathcal{G}_{CL}) < 0$ represents a decrease for functional $h_{\mathcal{G}}(\mathcal{G}_{CL})$,  denoting that cluster agents are driven close to the event origin and among themselves at the highest possible level to maintain connectivity with the rest of the network. Indeed, for a large number of nodes in $\mathcal{G}_{CL}$, it is reasonable to assume\footnote{Moving just one agent in cluster $\mathcal{G}_{CL}$ do not change graph volumes $\epsilon_{S}$, $\epsilon_{\overline{S}}$ significantly, if a large number of agents has already been deployed.} 
that $|\Delta \epsilon_{S}| \ll \epsilon_{S}$ and $|\Delta \epsilon_{\overline{S}}| \ll \epsilon_{\overline{S}}$. Therefore, trying to preserve the previously established topology without removing edges, it would hold that $\Delta h_{\mathcal{G}}(\mathcal{G}_{CL}) \simeq \Delta\epsilon_{C}(\epsilon_{S}^{-1}+\epsilon_{\overline{S}}^{-1}) > 0$. Whereas, in the setting where $h_{\mathcal{G}}(\mathcal{G}_{CL}) := h_{CL} $ and $\Delta \epsilon_{C}:=0$, it is possible -- by showing that $\Delta\epsilon_{S}  +\epsilon_{S} > 0 $ -- to conclude that $\Delta h_{\mathcal{G}}(\mathcal{G}_{CL}) = - \epsilon_{C}\Delta \epsilon_{S} / \left[\epsilon_{S}(\epsilon_{S}+\Delta\epsilon_{S})\right] < 0$ if and only if $\Delta\epsilon_{S} >0$ (line \ref{line:local_min_cond}). Denoting with $\Delta \mathcal{N}_j(t+\Delta t) := \left\lbrace v_{\bar{k}} \in \mathcal{G}_{CL} ~|~ e_{j\bar{k}}(t+\Delta t) \in \mathcal{E}\right\rbrace$ the set of new vertices that node $v_j$ can acquire as neighbors moving in $\mathbf{p}_j(t+\Delta t)$, s.t. $\Delta t \in [0,T]$, and with $\mathcal{N}_j(t+\Delta t) := \mathcal{N}_j(t) \cup \Delta \mathcal{N}_j(t+\Delta t)$ the new neighborhood for $v_j$, the $j$-th contribution to the vertex volume variation of cluster $\mathcal{G}_{CL}$ is yielded by
\begin{align}\label{eq:volume_variation}
\Delta_j \mathrm{vol}_{v}(\mathcal{G}) =& \sum\limits_{\forall v_{\bar{k}}\in \Delta \mathcal{N}_j(t+\Delta t)} |v_{\bar{k}}|  - \left|\mathcal{N}_{j}(t) \cap \mathcal{G}_{CL}\right| |v_{j}(t)| +\nonumber \\
& \left|\mathcal{N}_j(t+\Delta t) \cap \mathcal{G}_{CL}\right| \hat{f}_{EV}(\mathbf{p}_{j}(t+\Delta t)).
\end{align}
Since each agent $a_{j}$ moves while all the others do not change their position, i.e. each $\mathbf{p}_{\bar{k}}$ remains constant for all $v_{\bar{k}} \in \Delta \mathcal{N}_{j}(t+\Delta t)$, it holds that $\Delta_j \mathrm{vol}_{v}(\mathcal{G}) = \Delta \mathrm{vol}_{v}(\mathcal{G}) = \Delta \epsilon_{S}$: this relation proves that Alg. \ref{alg:dispatch} is distributed. Moreover, since $\epsilon_{S} = \sum\limits_{\forall v_{\bar{k}} \in \mathcal{G}_{CL}} \left|v_{\bar{k}}(t)\right| \left| \mathcal{N}_{\bar{k}}(t) \cap \mathcal{G}_{CL}\right| > \left|\mathcal{N}_{j}(t) \cap \mathcal{G}_{CL}\right| |v_{j}(t)|$ is always verified, relation $\Delta\epsilon_{S} > -\epsilon_{S} $ holds true. In addition, to obtain an estimate of $\Delta_j \mathrm{vol}_{v}(\mathcal{G})$ whenever the environment is noisy, we decide to take into account the signal-to-noise ratio $SNR_{w}$ between the noise $w$ and estimates $\hat{f}_{EV}$. Dividing term $- \left|\mathcal{N}_{j}(t) \cap \mathcal{G}_{CL}\right| |v_{j}(t)|$ in \eqref{eq:volume_variation} by the quantity $(1+\alpha_{w}SNR_{w}^{-1})$, where $\alpha_{w} > 0$ is a tunable constant, facilitates the establishment of new communication links. Furthermore, whenever a dispatch session starts, node weights are updated (line \ref{line:filtering}), by means of a filtering procedure, e.g. by using a moving average FIR filter for each agent $a_{i}$, acting in the discrete time window where $a_{i}$ does not change position.\\
Once again, Alg. \ref{alg:dispatch} should be seen as a distributed one-hop loop, not as a recursive function because, thanks to the properties of $\mathcal{N}_{i}^{\star}$, only counters $c_{di}$ have to be stored and just incremented (line \ref{line:c_di}) for each node per each session. Lastly, variables $c_{d}^{\star}$, $\text{f}_{d}^{\star}$ are the only to be passed from node to node whenever a hop from $v_i$ to $v_j$ takes place.
\begin{algorithm}
	\caption{Focus on event: agents' dispatch\\ \textbf{function} $\left[c_{d}^{\star}, \text{f}_{d}^{\star}\right] =$\textsc{Dispatch}($v_i$,$c_{d}^{\star}$,$\text{f}_{d}^{\star}$)} \label{alg:dispatch}
	\begin{algorithmic}[1]
		\small
		\If{$c_{di} < c_{d}^{\star}$}
			\State $c_{di} \leftarrow c_{d}^{\star}$; \label{line:c_di}
		\EndIf
		\State $|v_{i}| \leftarrow $\textsc{Filtering}($|v_{i}|$,$\hat{f}_{EV}(\mathbf{p}_{i})$); \label{line:filtering}
		\State $\mathcal{N}_{i}^{\star} \leftarrow  (\mathcal{N}_{i} \cap \mathcal{G}_{CL}) \setminus \left\lbrace v_k\in \mathcal{N}_{i} ~|~ c_{dk}=c_{d}^{\star}\right\rbrace$; \label{line:restricted_neigh}
		\For{$k_{i}=1,...,|\mathcal{N}_{i}^{\star}|$} \label{line:k_i}
			\State $v_j \leftarrow \left[\mathcal{N}_{i}^{\star}\right]_{k_i}$;
			\While{$a_j$ is moving from $\mathbf{p}_j(t)$ to $\mathbf{p}_j(t+T)$}
				\If{$v_j$ cannot take a step forward}
					\State \textbf{break while} \label{line:collision_visib_issues}
				\EndIf
				\If{$\Delta_j \mathrm{vol}_{v}(\mathcal{G}) > 0$} \label{line:local_min_cond}
					\State let $a_j$ move from $\mathbf{p}_{j}(t)$ to $\mathbf{p}_{j}(t+\Delta t)$;
					\State $|v_j| \leftarrow \hat{f}_{EV}(\mathbf{p}_j(t+\Delta t))$
					\State $\mathcal{N}_j(t+\Delta t) \leftarrow \mathcal{N}_j(t) \cup \Delta \mathcal{N}_j(t+\Delta t)$;
					\State $\text{f}_{d}^{\star} \leftarrow \mathbf{false}$;
				\Else
					\State \textbf{break while} \label{line: stop_cond_deltavol}
				\EndIf
			\EndWhile
			\State $\left[c_{d}^{\star}, \text{f}_{d}^{\star}\right] \leftarrow$\textsc{Dispatch}($v_j$,$c_{d}^{\star}$,$\text{f}_{d}^{\star}$);
		\EndFor	
	\end{algorithmic}
\end{algorithm}

\section{On the number of deployed agents} \label{sec:number_deployed_agents}
Given the lack of metric information, knowing exactly beforehand how many agents are deployed in the coverage stage is generally arduous; thus, we have decided just to provide bounds in a simplified framework. In particular, we start considering the case in which the scenario $SC$ is an obstacle-free rectangle (A1) with dimensions $b_{SC} \times h_{SC}$; then, a further extension will be proposed. In addition, we also suppose that $r_{v} \geq 4 r_{b} $ to allow a proper hexagonal packing policy implementation (A2), as far as it is possible. With these ideas in mind, we state the following basic but crucial
\begin{Prop}\label{prop1}
\textit{ Assume (A1) and (A2) hold. Let us define the dimensionless quantities $\varrho_{b} := b_{SC}/r_{v}$, $\varrho_{b3} := \varrho_{b}/\sqrt{3}$, $\varrho_{h} := h_{SC}/r_{v}$, $\varrho_{h3} := \varrho_{h}/\sqrt{3}$ and the real scalar functions in the positive variables $(\varrho_{1},\varrho_{2})$
	\begin{align}
	\overline{g_{SC}}(\varrho_{1},\varrho_{2}) := & 1+\lfloor\varrho_{1}\rfloor \lfloor\varrho_{2}\rfloor + \lceil\varrho_{1}\rceil \lceil\varrho_{2}-1/2\rceil; \label{eq:ugsct} \\
	g_{SC}(\varrho_{1},\varrho_{2}) :=& 1+\ceil[\bigg]{\varrho_{1}-\sqrt{4-\varrho_{2}^2}} ; \\
	\underline{g_{SC}}(\varrho_{1},\varrho_{2}) :=& \lceil\varrho_{1}\rceil (\lceil\varrho_{2}+1/2\rceil + \lfloor\varrho_{2}-1\rfloor)+ \nonumber\\  &+\lfloor\varrho_{1}\rfloor \lceil\varrho_{2}-1\rceil - \lfloor\varrho_{1}+1\rfloor \lfloor\varrho_{2}\rfloor. \label{eq:lgsct}
	\end{align}
	For $\varrho_{b3} > 1$ and $ \varrho_{h3} > 1$, the minimum number of deployed agents $n_{a}$ to attain a complete coverage of the scenario can be upper bounded by $ \overline{n_{ag}}$ and lower bounded by $ \underline{n_{ag}}$, such that
	\begin{align}
	\overline{n_{ag}} &:= \min(\overline{g_{SC}}(\varrho_{b},\varrho_{h3}),\overline{g_{SC}}(\varrho_{h},\varrho_{b3})); \label{eq:upperNa}\\
	\underline{n_{ag}} &:= \min(\underline{g_{SC}}(\varrho_{b},\varrho_{h3}),\underline{g_{SC}}(\varrho_{h},\varrho_{b3})).\label{eq:lowerNa}
	\end{align} 
	Moreover, it holds that
	\begin{equation}\label{eq:Na_rettlungoestretto}
	n_{a} = \begin{cases}
	g_{SC}(\varrho_{b},\varrho_{h}), \quad \varrho_{h3} \leq 1 ~\text{and}~ \varrho_{b} > 1;\\
	g_{SC}(\varrho_{h},\varrho_{b}), \quad \varrho_{b3} \leq 1 ~\text{and}~ \varrho_{h} > 1;\\
	1, \qquad\qquad~~~  \varrho_{b} \leq 1 ~\text{and}~ \varrho_{h} \leq 1.
	\end{cases}
	\end{equation}
}
\end{Prop}
\vspace{0.3cm}
\begin{proof}
	Whenever $\varrho_{b3} > 1 ~\text{and}~ \varrho_{h3}>1$ holds, it is possible to cover a rectangular surface as shown in Fig. \ref{fig:bounds}.
	\begin{figure}[t!]
		\centering
		\includegraphics[width=0.45\textwidth, height=0.18\textwidth]{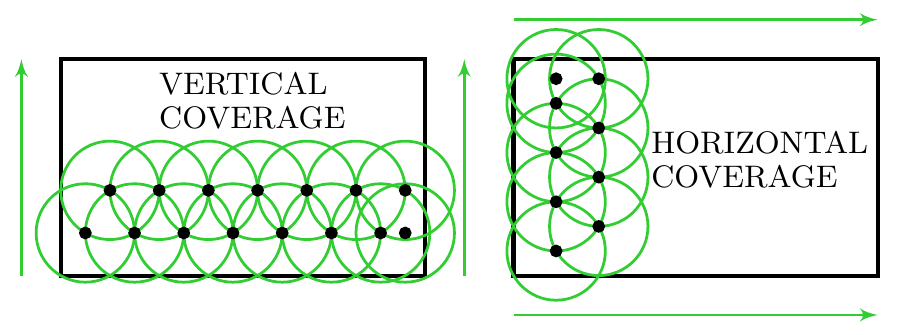}
		\caption{Vertical and horizontal coverage policies.}
		\label{fig:bounds}
	\end{figure}
	The choice of an appropriate coverage policy allows to determine an upper and a lower bound for the number of deployed agents $n_{a}$, in this case scenario. For this reason, in order to compute $\overline{n_{ag}}$ and $\underline{n_{ag}}$ in \eqref{eq:upperNa} and \eqref{eq:lowerNa} respectively, it is  necessary to minimize among $\overline{g_{SC}} (\varrho_{1},\varrho_{2})$ and $\underline{g_{SC}} (\varrho_{1},\varrho_{2})$, selecting the most convenient configuration among those shown in Fig. \ref{fig:bounds}.\\
	Let us define the fractional part of a real number $\omega \in \mathbb{R}$ as $\mathrm{frac}(\omega) := \omega - \lfloor \omega \rfloor$, for $\omega \geq 0$, and the characteristic function $\chi_{\Omega}(\omega)$ on set $\Omega$, $1$-valued if $\omega \in \Omega$, $0$-valued otherwise.
	W.l.o.g., let us adopt the vertical coverage policy in Fig. \ref{fig:bounds}. Then, $\underline{g_{SC}}(\varrho_{b},\varrho_{h3}) \leq n_{a}$ holds by assigning
	\begin{align}
		\underline{g_{SC}}(\varrho_{b},\varrho_{h3}) := \left(\lfloor \varrho_{b} \rfloor +\lceil \varrho_{b} \rceil \right)\lfloor \varrho_{h3} \rfloor + \nonumber \\ \lceil\varrho_{b}\rceil \chi_{\left\lbrace \omega | \omega > 1/2\right\rbrace} \left( \mathrm{frac}(\rho_{h3})\right)+ \nonumber \\
		-\lfloor \varrho_{h3} \rfloor \chi_{\left\lbrace 0\right\rbrace}\left( \mathrm{frac}\left(\rho_{b} \right) \right)  + \nonumber \\
		-\lfloor \varrho_{b} \rfloor \chi_{\left\lbrace 0\right\rbrace}\left( \mathrm{frac}\left(\rho_{h3} \right) \right). \label{eq:lgsc}
	\end{align}
	In \eqref{eq:lgsc}, first and second terms represent the space occupancy for agents at the first and second rows in Fig. \ref{fig:bounds}, while third and fourth terms take into account redundant agents placed on the boundaries. Similarly, $\overline{g_{SC}}(\varrho_{b},\varrho_{h3}) \geq n_{a}$ holds by assigning
	\begin{align}
		\overline{g_{SC}}(\varrho_{b},\varrho_{h3}) := 1+\left(\lfloor \varrho_{b} \rfloor +\lceil \varrho_{b} \rceil \right)\lfloor \varrho_{h3} \rfloor + \nonumber \\ \lceil\varrho_{b}\rceil \chi_{\left\lbrace \omega | \omega > 1/2\right\rbrace} \left( \mathrm{frac}(\rho_{h3})\right). \label{eq:ugsc} 
	\end{align}
	Expressions \eqref{eq:ugsct} and \eqref{eq:lgsct} are then obtained by leveraging floor and ceiling properties starting from \eqref{eq:ugsc} and \eqref{eq:lgsc}, respectively. Now, w.l.o.g., let us assume that the width $b_{SC}$ is larger than the height $h_{SC}$. If $\varrho_{h3} \leq 1 ~\text{and}~ \rho_{b} >1$ holds then the scenario surface can be exactly covered with the following number of agents:
	\begin{equation}\label{eq:Nagsc}
	n_{a} = g_{SC}(\varrho_{b},\varrho_{h}) := 1+ \ceil[\Big]{\varrho_{b}-2\sqrt{1-(\varrho_{h}/2)^2}} .
	\end{equation}
	Equality in \eqref{eq:Nagsc} holds true only if $\varrho_{b}-2\sqrt{1-(\varrho_{h}/2)^2} > -1$ is satisfied; however, the latter relation is already verified, since condition $\varrho_{h3} \leq 1 ~\text{and}~ \rho_{b} >1$ characterizes this case scenario. Finally, for $\varrho_{b} \leq 1 ~\text{and}~ \varrho_{h} \leq 1$, one agent is trivially sufficient to cover the entire surface.
\end{proof}
This result, can be further extended by taking into account border effects while the algorithm compute deployment positions. For instance, the fact that a rectangular scenario SC has a well defined perimeter $p_{SC} = 2(b_{SC}+h_{SC})$ suggests us that a good heuristic to approximatively improve lower bound in \eqref{eq:lowerNa} can be adopted by adding to $\underline{n_{ag}}$ the quantity $\underline{n_{ap}} = \lfloor p_{SC}/r_{v} \rfloor$. Finally, an additional step can be taken in order to provide a rough estimate of the lower bound $\underline{n_{ag}}(SC)$ for a generic scenario $SC$. Since any connected space in $\mathbb{R}^{2}$ can be easily approximated by a segmentation into rectangles $\left\lbrace re_{k}\right\rbrace_{k=1}^{l_{re}}$ , it holds that 
\begin{align}\label{eq:lbound_generic}
\underline{n_{ag}}(SC) \simeq  \sigma_{p0}\underline{n_{ap}}(EN)+ \sum\limits_{k=1}^{l_{ob}}  \sigma_{pk}\underline{n_{ap}}(ob_{k}) + \nonumber \\
+ \sum\limits_{k=1}^{l_{re}} \sigma_{gk}\underline{n_{ag}}(re_{k}^{2}) 
\end{align}
where the dependency of each bound on a precise element of the scenario is indicated inside round brackets. In \eqref{eq:lbound_generic}, coefficients $\sigma_{gk}$, for $k=1,...,l_{re}$, are either equal to $1$, if $ re_{k}^{2} \subseteq CS^{2}$ or $0$, otherwise and coefficients $\sigma_{pk}\in [0,1]$, for $k=0,...,l_{ob}$, are selected to describe how much a delimitation conveys border effects.

\columnbreak
\begin{figure*}[h!] 
	\centering
	\subfigure[Coverege and cluster selection]{\includegraphics[height=0.27\textwidth, trim={0.8cm 3.5cm 2.1cm 4.5cm},clip ]{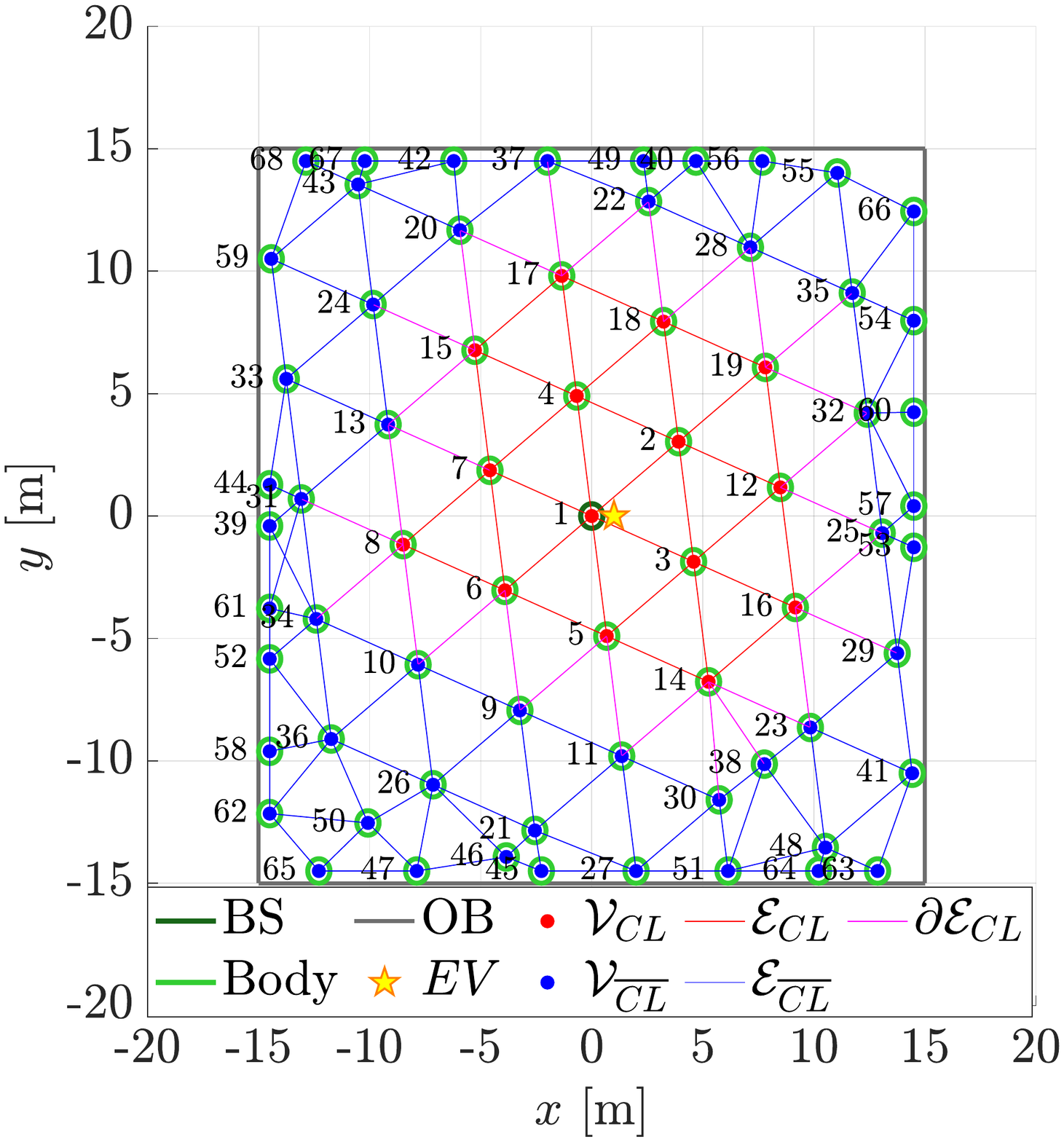}\label{fig:r11}}
	\subfigure[Event intensity]{\includegraphics[height=0.26\textwidth, trim={0.8cm 3.5cm 2.3cm 3.5cm},clip]{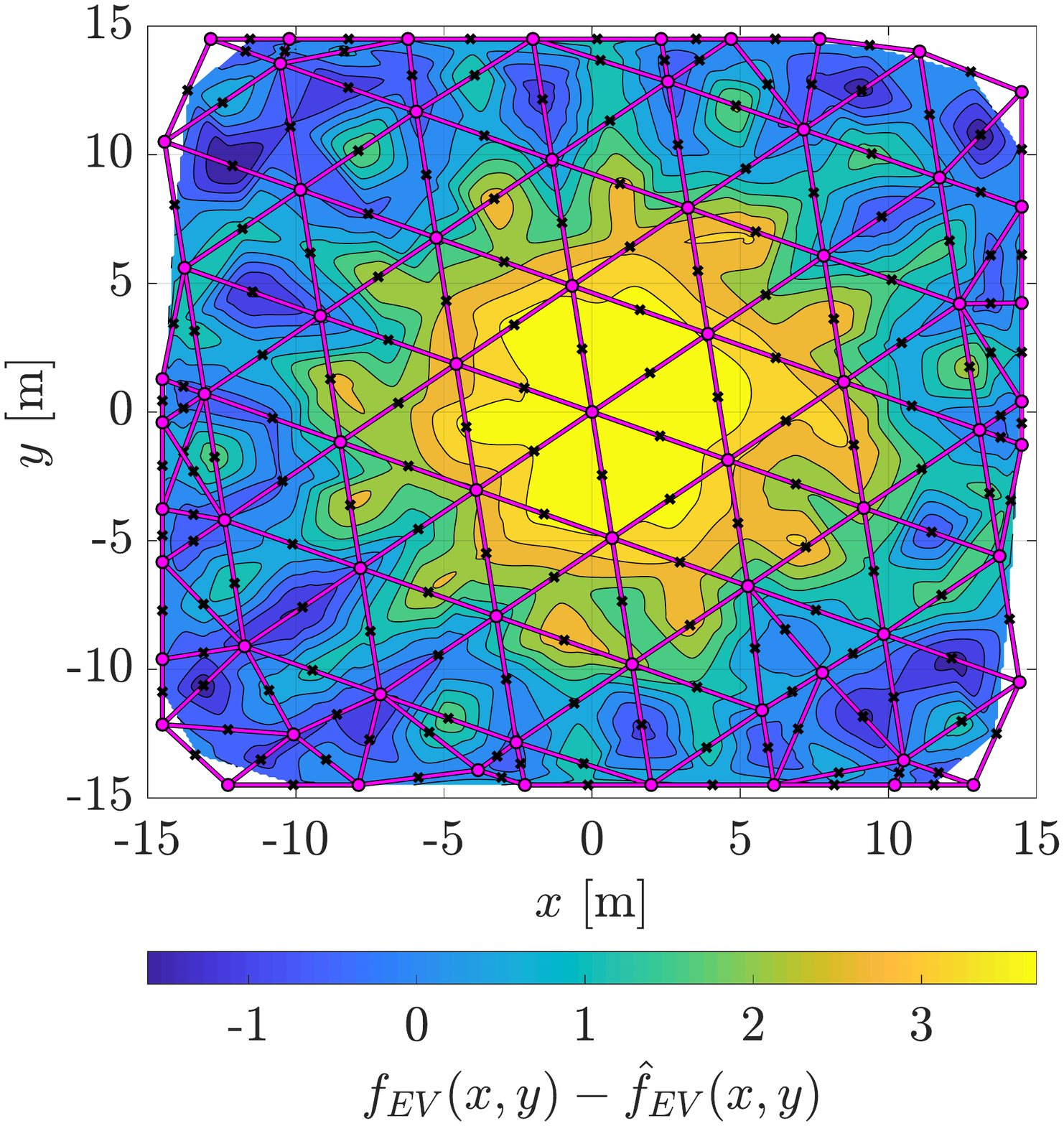}\label{fig:e11}}
	\subfigure[Dispatch of the selected cluster]{\includegraphics[height=0.26\textwidth, trim={0.8cm 3.5cm 2.1cm 4.5cm},clip]{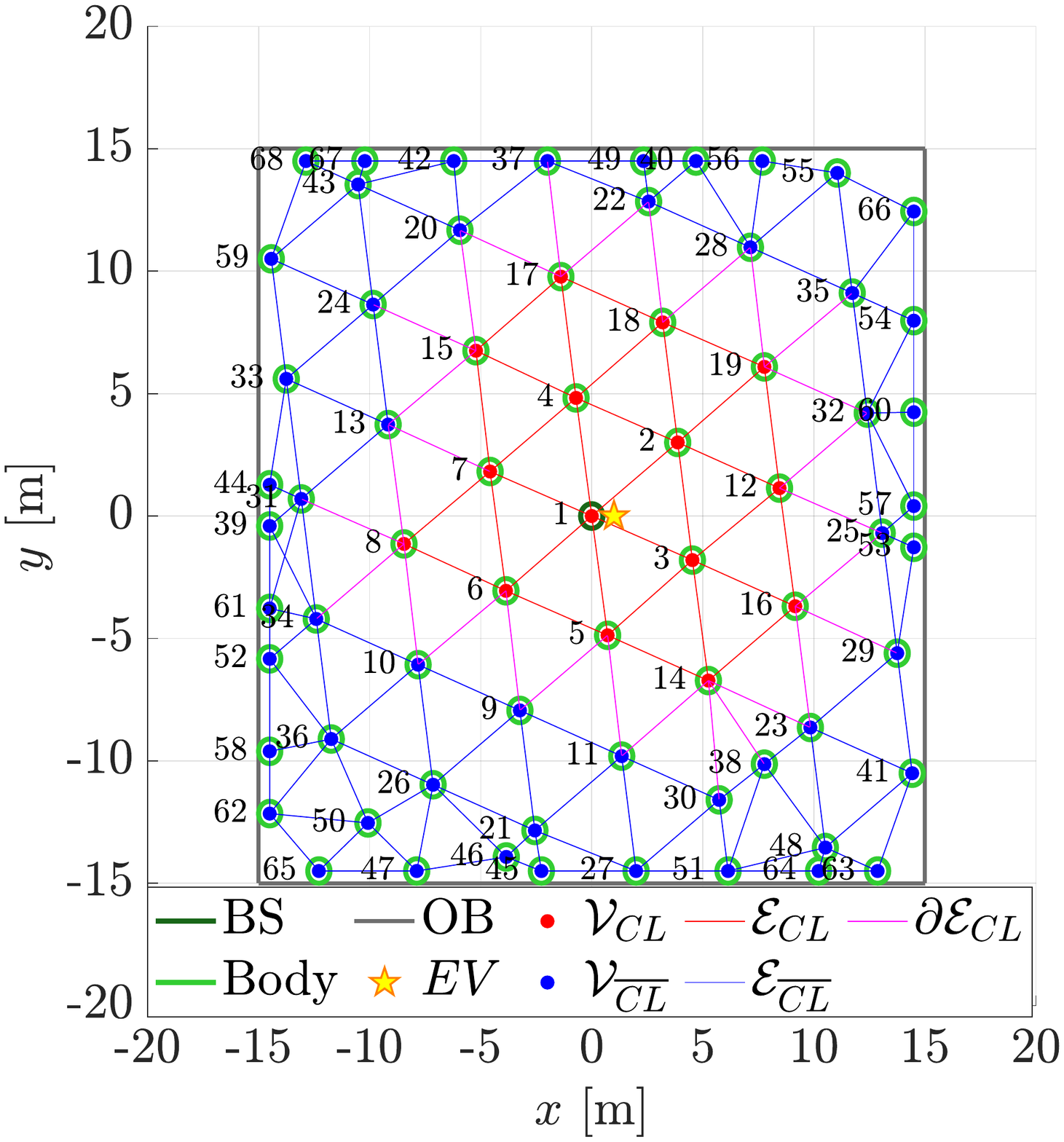}\label{fig:p11}}
	\subfigure[Isoperimetric functional]{\includegraphics[height=0.26\textwidth, trim={ 0.1cm 3cm 1.8cm 3.1cm},clip]{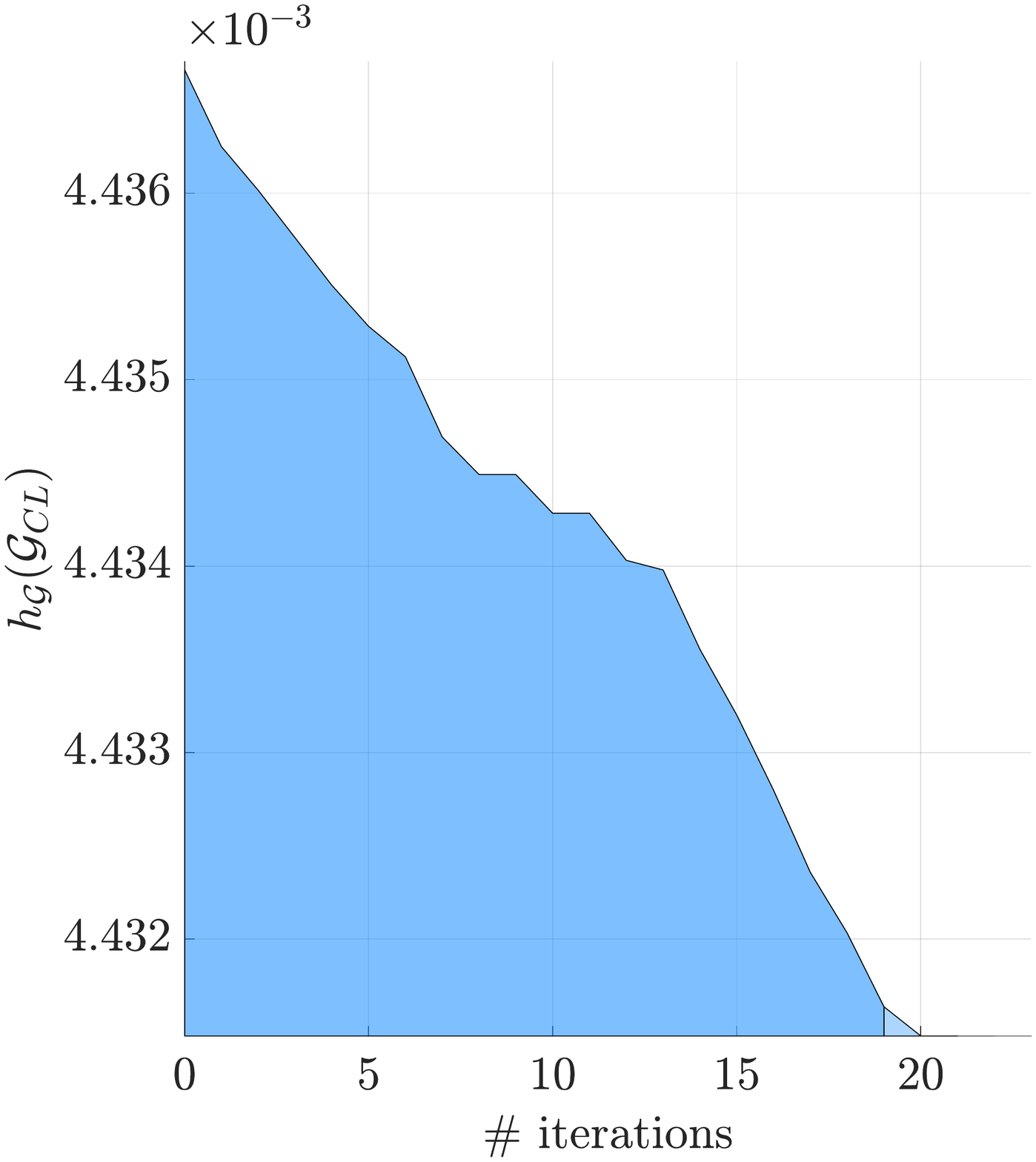}\label{fig:f11}}
	\vspace{-0.2cm}
	\caption{Dynamic coverage in an obstacle-free scenario. Hexagonal packing is mostly achieved, up to border effects, and a cluster (red dots) is selected around the event source (yellow star). The agent dispatch has practically no effect as expected, since topology cannot shrink towards the event. After $20$ iterations and $2$ sessions, the execution is terminated.}
	\label{fig:1}
\end{figure*} 
\vspace{0.1cm}
\columnbreak
\begin{figure*}[h!]
	\vspace{-0.2cm}
	\centering
	\subfigure[Coverege and cluster selection]{\includegraphics[height=0.26\textwidth, trim={ 0.8cm 3.5cm 2.1cm 4.5cm},clip]{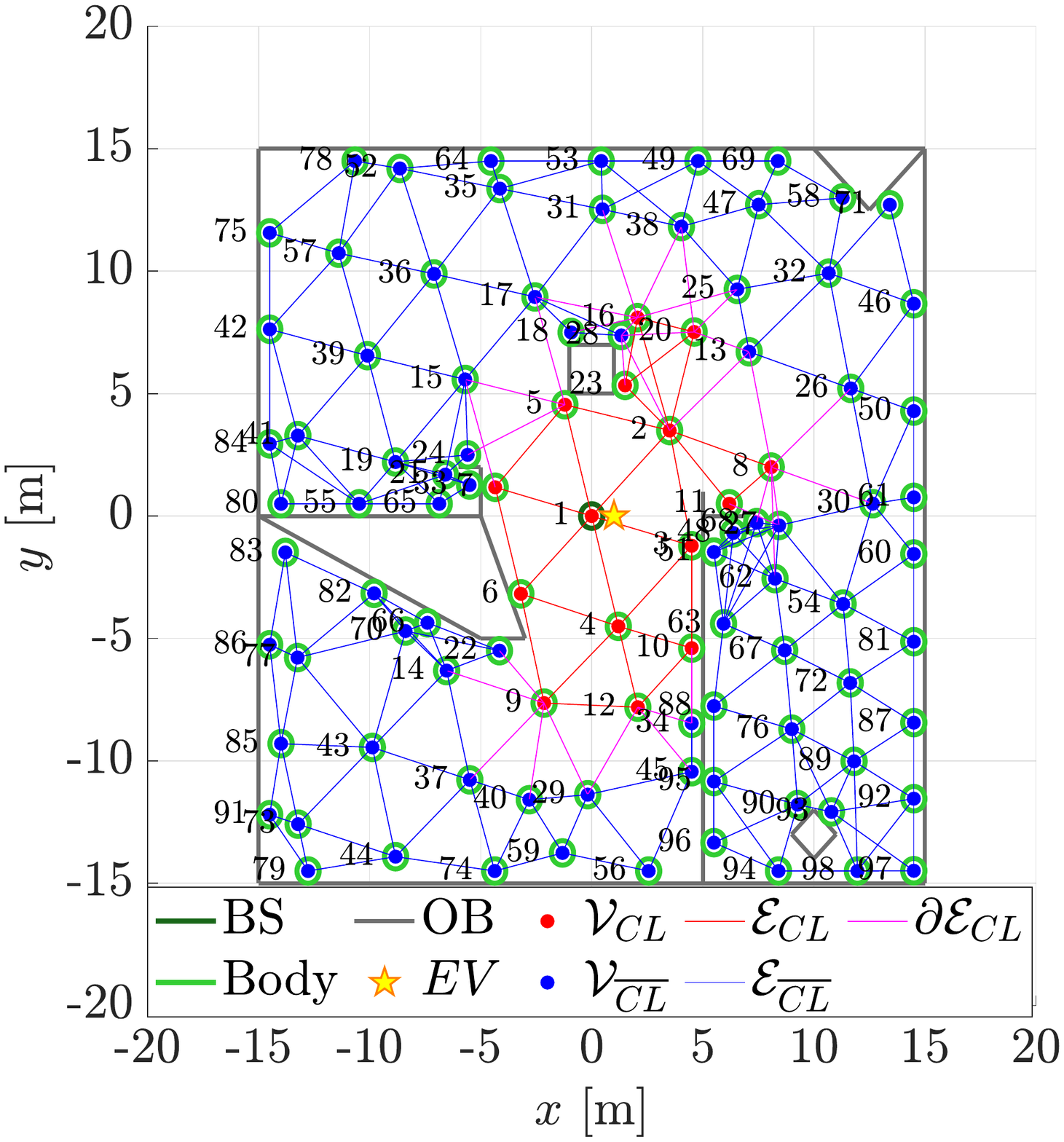}\label{fig:r21}}
	\subfigure[Event intensity]{\includegraphics[height=0.26\textwidth, trim={0.8cm 3.5cm 2.3cm 3.5cm},clip]{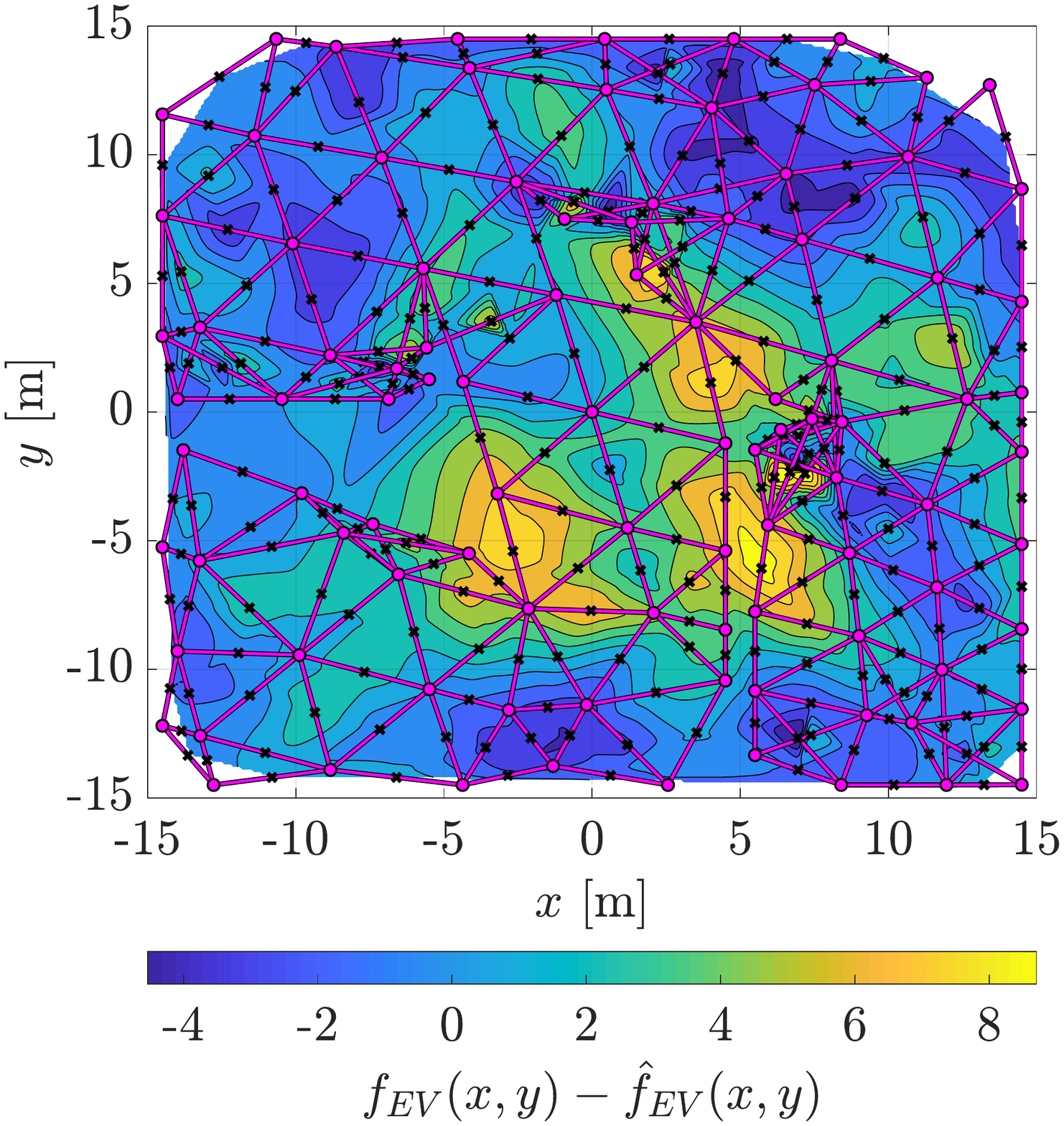}\label{fig:e21}}
	\subfigure[Dispatch of the selected cluster]{\includegraphics[height=0.26\textwidth, trim={ 0.8cm 3.5cm 2.1cm 4.5cm},clip]{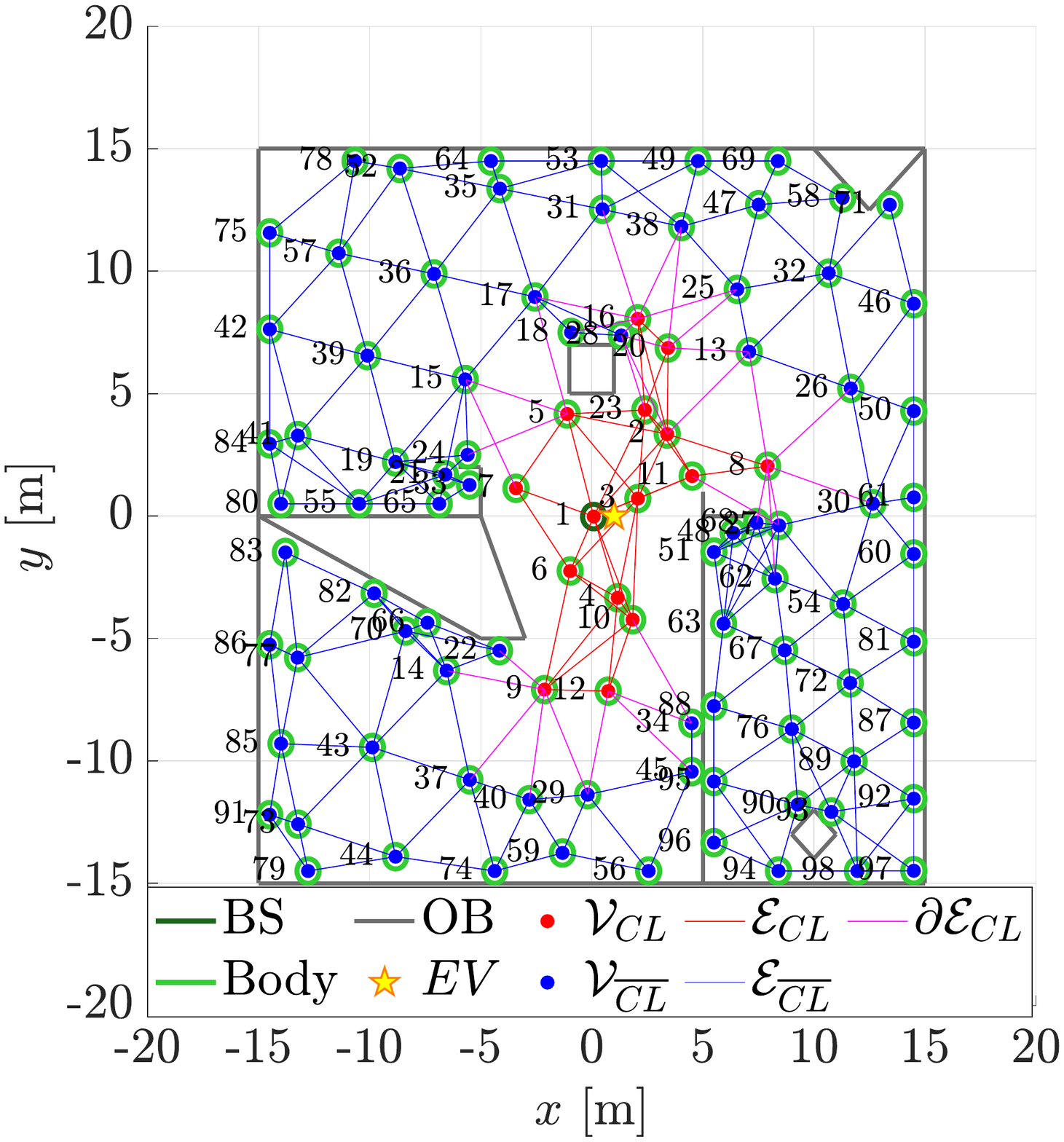}\label{fig:p21}}
	\subfigure[Isoperimetric functional]{\includegraphics[height=0.26\textwidth, trim={0.1cm 3cm 1.8cm 3.1cm},clip]{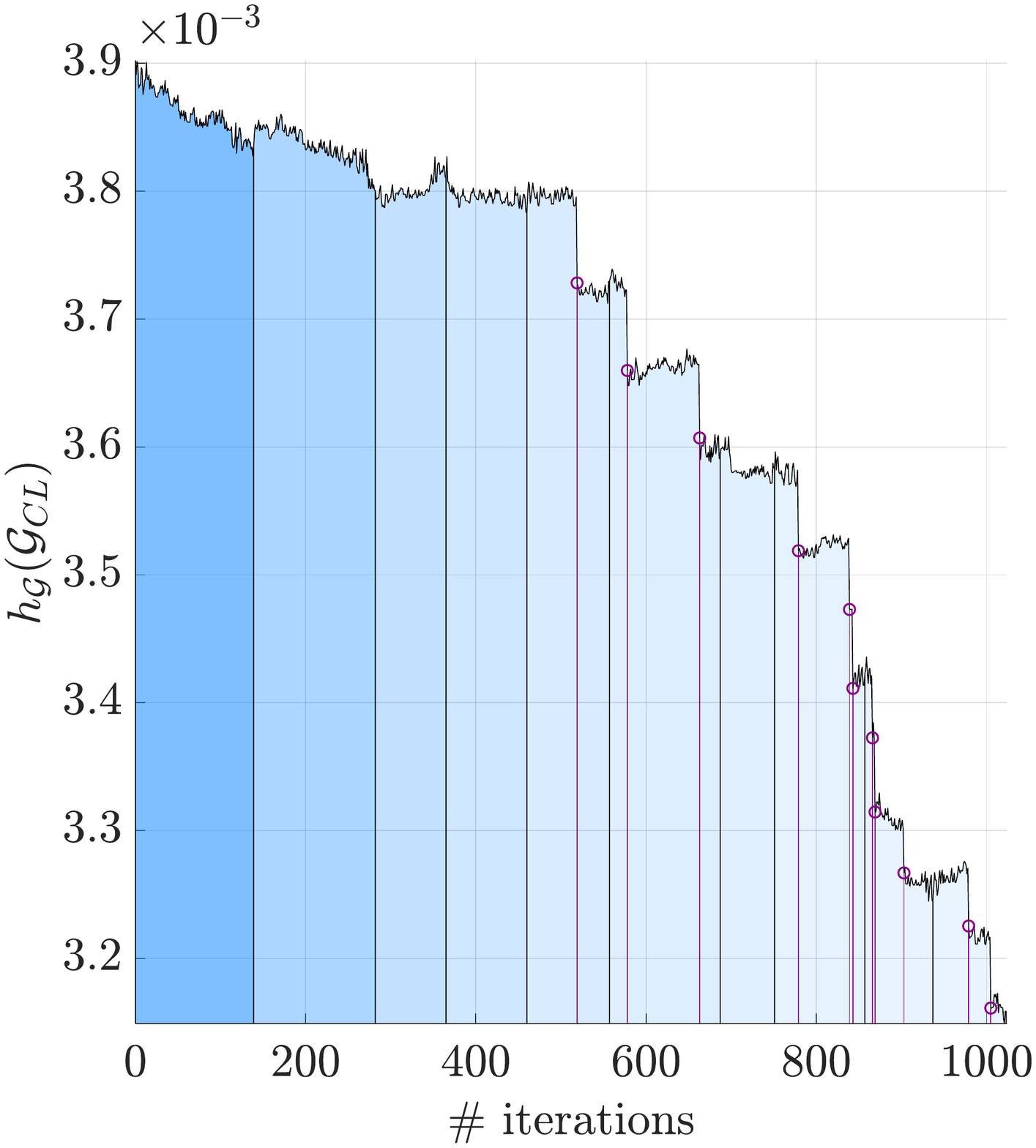}\label{fig:f21}}
	\vspace{-0.2cm}
	\caption{Dynamic coverage in a noisy structured scenario. Hexagonal packing is achieved only in few areas, because of the large presence of obstacle borders. The agent dispatch leads to a graph topology narrowing by increasing the cluster volume: whenever an edge between two nodes in the cluster is added, the isoperimetric functional decreases with a discontinuity (purple spikes). Execution stops after $10$ sessions, with more than $1000$ iterations.}
	\label{fig:2}
\end{figure*} 

\begin{figure*}[h!]
\centering
\subfigure[Dispatch of the selected cluster]{\includegraphics[height=0.26\textwidth, trim={0.8cm 3.5cm 2.1cm 4.5cm},clip]{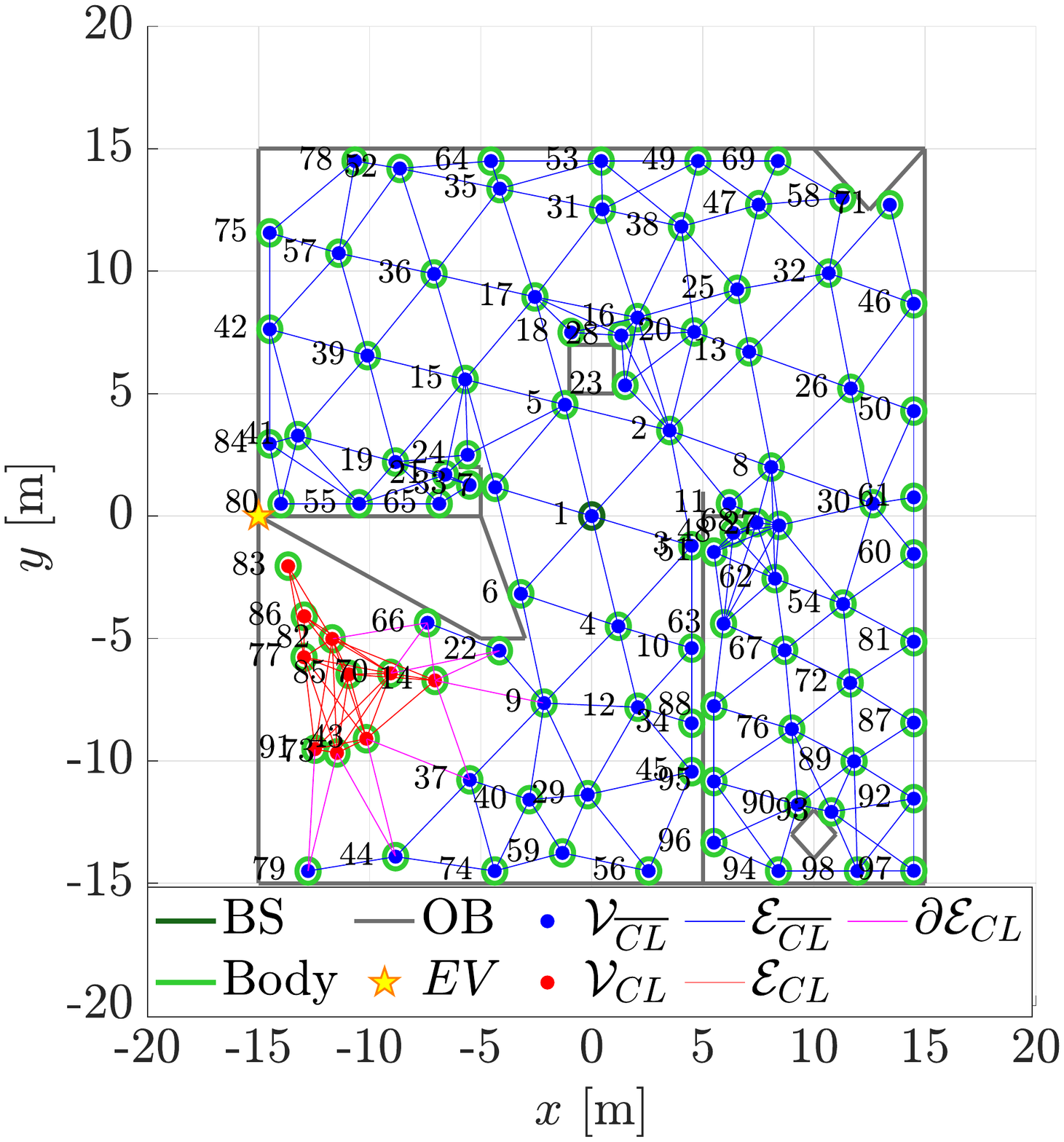}\label{fig:p23}}
\subfigure[Isoperimetric functional]{\includegraphics[height=0.26\textwidth, trim={0.1cm 3cm 1.8cm 3.1cm},clip]{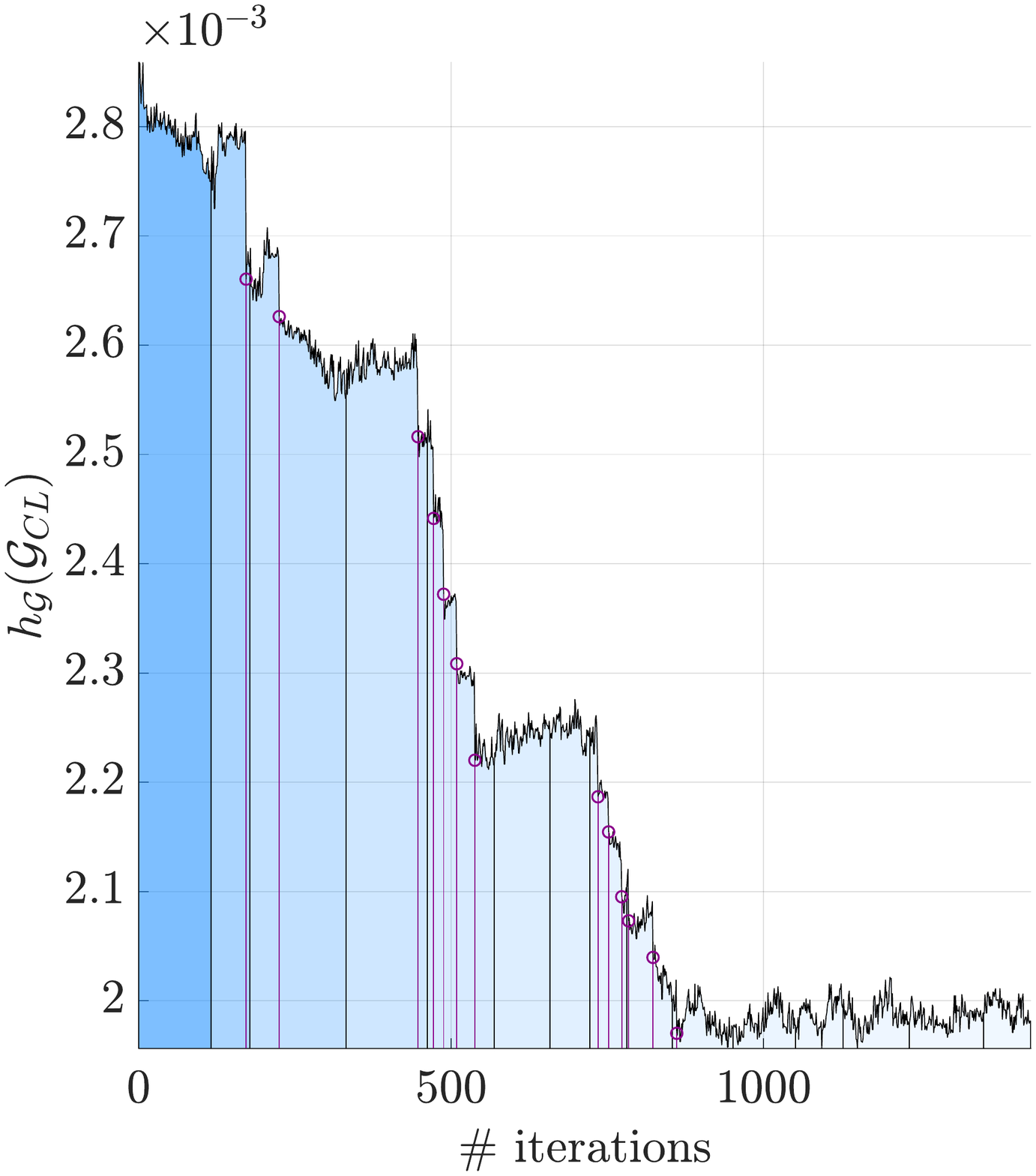}\label{fig:f23}}
\subfigure[Dispatch of the selected cluster]{\includegraphics[height=0.26\textwidth, trim={0.8cm 3.5cm 2.1cm 4.5cm},clip]{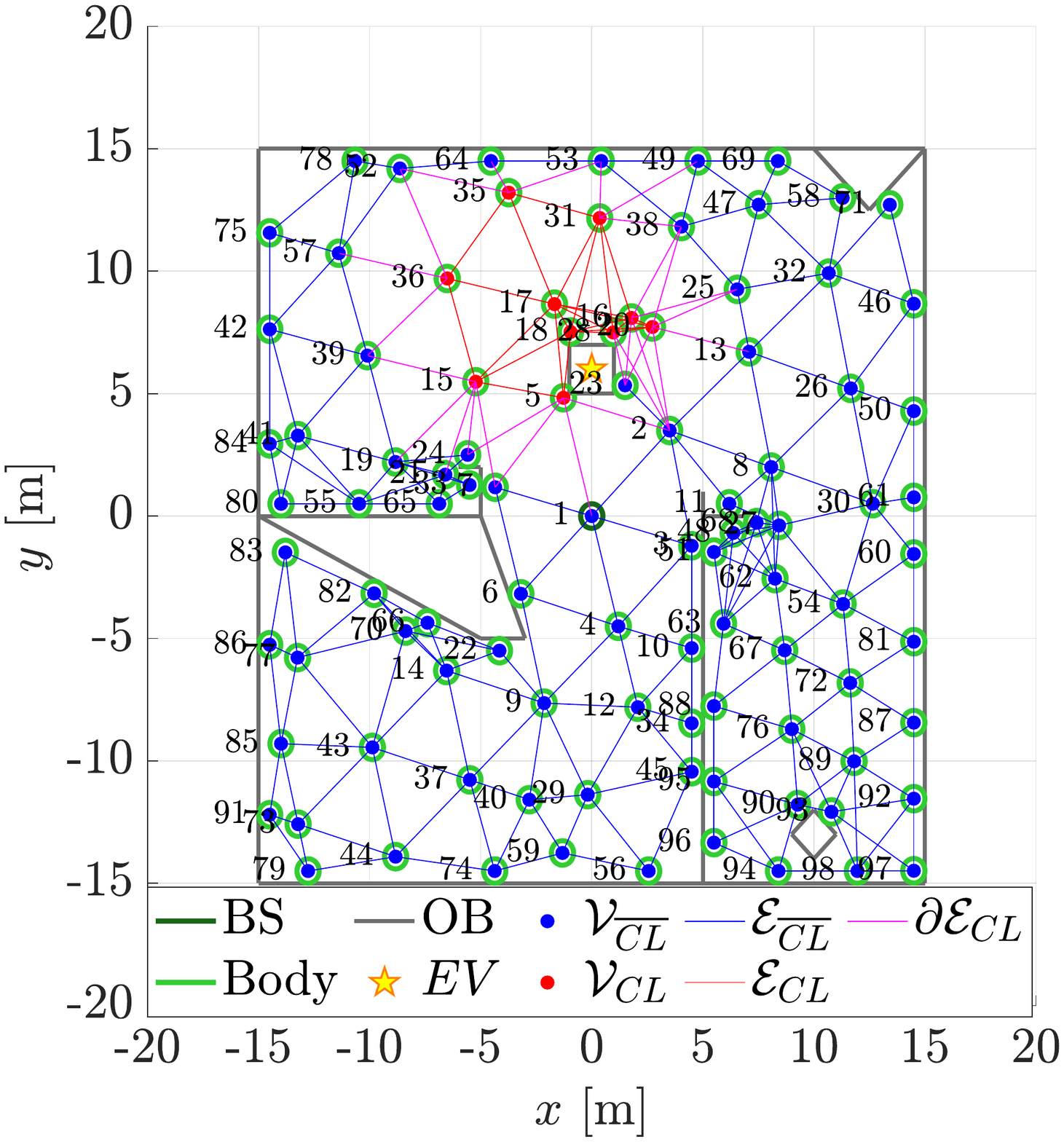}\label{fig:p33}}
\subfigure[Isoperimetric functional]{\includegraphics[height=0.26\textwidth, trim={0.1cm 3cm 1.8cm 3.1cm},clip]{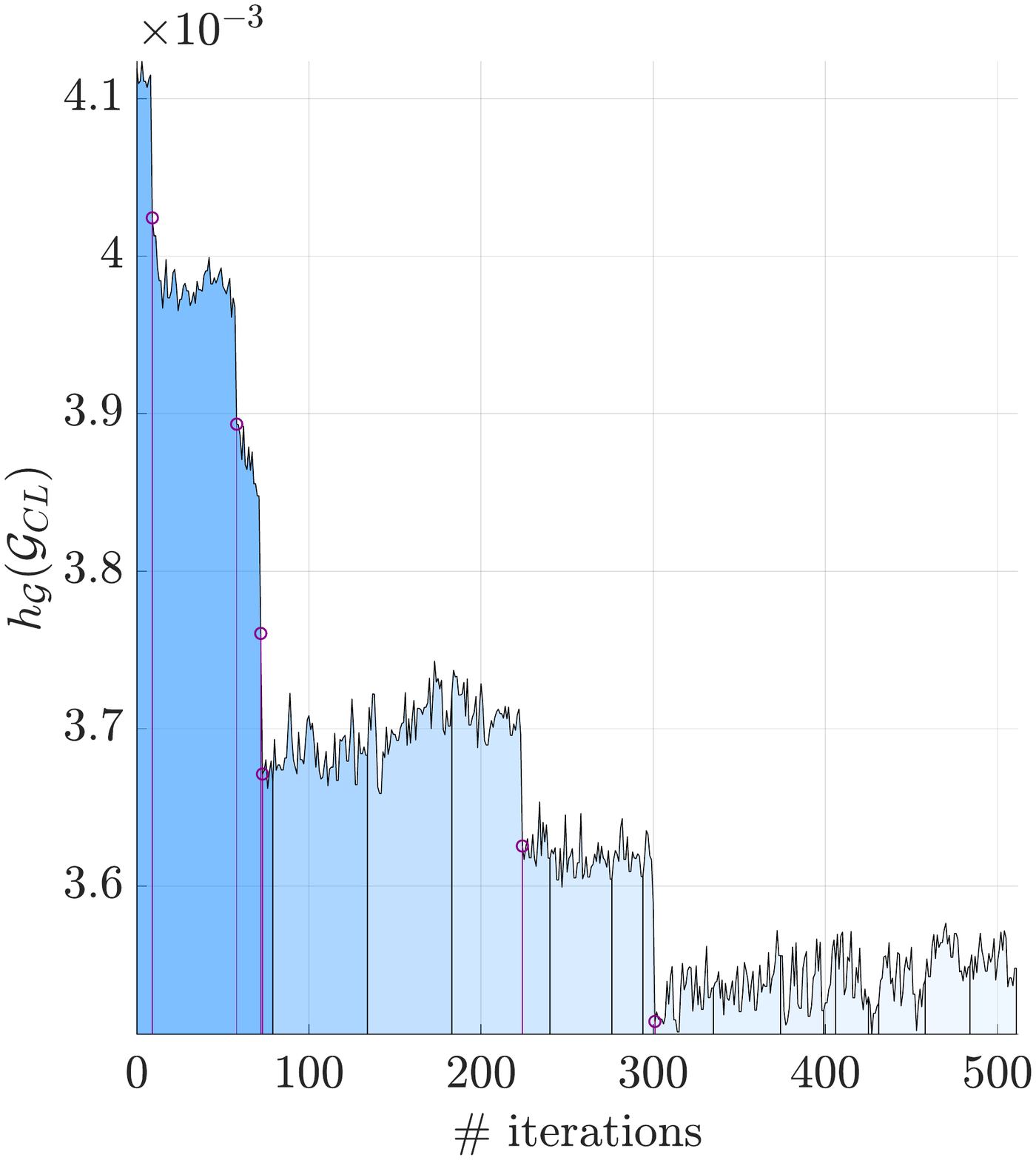}\label{fig:f33}}
\caption{Dynamic coverage in a highly noisy structured scenario with unaccessible event sources. In (a) and (b) the dispatch is attained towards an event placed on the enclosure. In (c) and (d) the dispatch is attained around squared obstacle. The abundant presence of noise limits the cluster selection and the poor sensing affects the dispatch performances negatively.}
\label{fig:3}
\end{figure*}

\section{Numerical simulations}\label{sec:numerical_simulations}
We demonstrate the validity and describe limitations of our algorithm providing numerical simulations in
\begin{itemize}
	\item A) an obstacle-free scenario $SC=EN$;
	\item B) structured environment $SC=(EN,OB)$ with presence of obstacles.
\end{itemize}
In this set-up, the following quantities are assigned: $r_{v}=5\si{\meter}$, $N_{T}=1$, $k_{EV}=160$, $r_{EV} = 15 \si{\meter}$.
Impacts against barriers, collision between agents and general dynamics are virtually implemented in a realistic fashion for all instances of the algorithm. We assume estimates $\hat{f}_{EV}(\mathbf{p}) = f_{EV}(\mathbf{p}) + w(f_{EV}(\mathbf{p}))$ are affected by a uniformly distributed noise $w(f_{EV}(\mathbf{p}))$ with zero mean and variance $f_{EV}(\mathbf{p})^{2}\sigma_{w}^{2}/3$; thus, $\alpha_{w} = 3$ is set.
\subsection{Obstacle-free scenario}
In this simulation, we test our algorithm in a $15\si{\meter} \times 15\si{\meter}$ squared scenario as Fig. \ref{fig:r11} shows. Neither barriers nor noise are present. A coverage with $n_{a}=68$ agents 
is attained. Cluster formation takes place thanks to event sensing, illustrated in Fig. \ref{fig:e11}, after a greedy selection of $n_{CL}=15$ members. Once the dispatch is completed, the spacial distribution of the agents does not practically change (Fig. \ref{fig:p11}), since the event $EV = (1\si{\meter},0)$ is situated in the middle of a quasi-perfect hexagonal packing. Finally, Fig. \ref{fig:f11} exhibits a slight decrease of functional $h_{\mathcal{G}}(\mathcal{G}_{CL})$: this phenomenon is due to borders effects in the scenario that influence agents located in a marginal position w.r.t. the scenario centroid.
\vspace{-0.2cm}
\subsection{Structured and noisy environment}
In Fig. \ref{fig:r21}, a simulation in a structured squared scenario is shown choosing the event in $EV = (1\si{\meter},0)$. It is remarkable to note that more agents w.r.t. the obstacle-free case have been deployed ($n_{a}=98$
). Cluster formation ($n_{CL}=15$ is set) mainly arises where communication links are concentrated and around the event source after the sensing phase, illustrated in Fig. \ref{fig:e21}. Differently from the previous case, during the dispatch stage (Fig. \ref{fig:p11} vs. Fig. \ref{fig:p21}), agents focus on the event and, as a result, the graph topology shrinks around and towards the point $EV$. Remarkably, many additional edges are added in cluster $\mathcal{G}_{CL}$: this fact can be observed in Fig. \ref{fig:f21}, where, in correspondence to each sharp decrease of functional $h_{\mathcal{G}}(\mathcal{G}_{CL})$, a new link between two cluster nodes is created. Moreover, another peculiarity is highlighted: the minimization of $h_{\mathcal{G}}(\mathcal{G}_{CL})$ requires several sessions and much more iterations to be accomplished w.r.t. to the obstacle-free case. This fact is also due to the presence of noise ($\sigma_{z}=0.01$ is set): the computation of volume variation $\Delta \mathrm{vol}_{v}(\mathcal{G}) $ is affected by uncertainty; therefore, wrong descent directions for $h_{\mathcal{G}}(\mathcal{G}_{CL})$ are selected during dispatch. We finally highlight that agent number $28$, as few other relevant agents close to the event, is not involved in the cluster because of the noisy environment, representing a potential limitation for our approach.

	\indent As a further assessment, the algorithm remains robust enough when events are selected in unreachable points of the scenario, e.g. $EV=(-15\si{\meter},0)$ on the enclosure (Fig. \ref{fig:p23}) and $EV=(0,6\si{\meter})$ inside an inaccessible obstacle (Fig. \ref{fig:p33}), even though these two event sources are arduous or impossible to be covered. In both cases, especially when the event is located on the enclosure, the isoperimetric functional $h_{\mathcal{G}}(\mathcal{G}_{CL})$ requires a large number of iterations and potentially infinite sessions to be minimized, as shown in Figs. \ref{fig:f23} and \ref{fig:f33}, but a focus on the event is eventually achieved, to some extent. However, in these more critical cases, few limitations emerge: the considerable presence of noise ($\sigma_{z} = 0.1$) and the lower desired number of agents selected ($n_{CL} = 10$) affect both clustering and dispatch performances negatively. For instance, cluster in Fig. \ref{fig:p33} does not involve agent number $23$; furthermore, in Fig. \ref{fig:p23} the dispatch drives agents fairly far from the event.

\vspace{-0.1cm}
\section{Conclusions and future directions} \label{sec:conclusions}
An algorithm for dynamic coverage and focus on event has been designed. The agents employed for this task are provided by a bearing-based visual homing controller relying on limited sensing capabilities and local information. Geometrical models can simulate space occupation in an unknown scenario admitting the presence of obstacles.\\
In the deployment stage, bounds for the number of deployed agents are given and 
a fully distributed implementation is proposed, as well as for cluster formations. Leveraging the minimization of an isoperimetric functional to increase the cluster volume and, consequently, maximizing the communication over the cluster, the dispatch of agents can be attained towards an event belonging to the environment.\\
Future work on complete dynamic coverage in a noisy framework is envisaged, since uncovered regions could arise while agents are steered.

\bibliographystyle{IEEEtran}
\bibliography{IEEEfull,biblio}

\begin{thebibliography}{10}
\providecommand{\url}[1]{#1}
\csname url@rmstyle\endcsname
\providecommand{\newblock}{\relax}
\providecommand{\bibinfo}[2]{#2}
\providecommand\BIBentrySTDinterwordspacing{\spaceskip=0pt\relax}
\providecommand\BIBentryALTinterwordstretchfactor{4}
\providecommand\BIBentryALTinterwordspacing{\spaceskip=\fontdimen2\font plus
\BIBentryALTinterwordstretchfactor\fontdimen3\font minus
  \fontdimen4\font\relax}
\providecommand\BIBforeignlanguage[2]{{%
\expandafter\ifx\csname l@#1\endcsname\relax
\typeout{** WARNING: IEEEtran.bst: No hyphenation pattern has been}%
\typeout{** loaded for the language `#1'. Using the pattern for}%
\typeout{** the default language instead.}%
\else
\language=\csname l@#1\endcsname
\fi
#2}}

\bibitem{MesbahiEgerstedt2010}
M.~Mesbahi and M.~Egerstedt, \emph{Graph Theoretic Methods in Multiagent
  Networks}.\hskip 1em plus 0.5em minus 0.4em\relax Princeton: Princeton
  University Press, 2010.

\bibitem{WangDjahelMcManis2014}
S.~Wang, S.~Djahel, and J.~McManis, ``A multi-agent based vehicles re-routing
  system for unexpected traffic congestion avoidance,'' in \emph{17th
  International IEEE Conference on Intelligent Transportation Systems (ITSC)},
  Oct 2014, pp. 2541--2548.

\bibitem{GaoXiaoLiuLiangChen2012}
J.~Gao, Y.~Xiao, J.~Liu, W.~Liang, and C.~P. Chen, ``A survey of
  communication/networking in smart grids,'' \emph{Future Generation Computer
  Systems}, vol.~28, no.~2, pp. 391 -- 404, 2012.

\bibitem{AlTurjman2018}
F.~Al-Turjman and S.~Alturjman, ``5g/iot-enabled uavs for multimedia delivery
  in industry-oriented applications,'' \emph{Multimedia Tools and
  Applications}, Jun 2018.

\bibitem{GeYangHan2017}
X.~Ge, F.~Yang, and Q.-L. Han, ``Distributed networked control systems: A brief
  overview,'' \emph{Information Sciences}, vol. 380, pp. 117 -- 131, 2017.

\bibitem{DuFaberGunzburger1999}
Q.~Du, V.~Faber, and M.~Gunzburger, ``Centroidal voronoi tessellations:
  Applications and algorithms,'' \emph{SIAM Review}, vol.~41, no.~4, pp.
  637--676, 1999.

\bibitem{RutishauserCorrellMartinoli2009}
S.~Rutishauser, N.~Correll, and A.~Martinoli, ``Collaborative coverage using a
  swarm of networked miniature robots,'' \emph{Robotics and Autonomous
  Systems}, vol.~57, no.~5, pp. 517 -- 525, 2009.

\bibitem{DudekJenkinMilosWilkes1991}
G.~Dudek, M.~Jenkin, E.~Milios, and D.~Wilkes, ``Robotic exploration as graph
  construction,'' \emph{IEEE Transactions on Robotics and Automation}, vol.~7,
  no.~6, pp. 859--865, Dec 1991.

\bibitem{Zomorodian2010}
A.~Zomorodian, ``Fast construction of the vietoris-rips complex,''
  \emph{Computers and Graphics}, vol.~34, no.~3, pp. 263 -- 271, 2010, shape
  Modelling International (SMI) Conference 2010.

\bibitem{SilvaGhrist2006}
V.~de~Silva and R.~Ghrist, ``Coordinate-free coverage in sensor networks with
  controlled boundaries via homology,'' \emph{The International Journal of
  Robotic Research}, vol.~25, no.~12, pp. 1205--1222, 2006.

\bibitem{GhristLipskyDerenickSperanzon2012}
R.~Ghrist, D.~Lipsky, J.~Derenick, and A.~Speranzon, ``Topological
  landmark-based navigation and mapping,'' technical report, 2012.

\bibitem{SilvaGhrist2007}
V.~de~Silva and R.~Ghrist, ``Coverage in sensor networks via persistent
  homology,'' \emph{Algebraic \& Geometric Topology}, vol.~7, pp. 339--358,
  2007.

\bibitem{RamaithitimaWhitzerBhattacharyaKumar2015}
R.~Ramaithitima, M.~Whitzer, S.~Bhattacharya, and V.~Kumar, ``Sensor coverage
  robot swarms using local sensing without metric information,'' in \emph{2015
  IEEE International Conference on Robotics and Automation (ICRA)}, May 2015,
  pp. 3408--3415.

\bibitem{WitteburgDziengelAdlerKasmi2012}
G.~Wittenburg, N.~Dziengel, S.~Adler, Z.~Kasmi, M.~Ziegert, and J.~Schiller,
  ``Cooperative event detection in wireless sensor networks,'' \emph{IEEE
  Communications Magazine}, vol.~50, no.~12, pp. 124--131, December 2012.

\bibitem{MehdiTayarani2014}
M.~M. Afsar and M.-H. Tayarani-N, ``Clustering in sensor networks: A literature
  survey,'' \emph{Journal of Network and Computer Applications}, vol.~46, pp.
  198 -- 226, 2014.

\bibitem{LukicStojmenovic2013}
M.~Lukic and I.~Stojmenovic, ``Energy-balanced matching and sequence dispatch
  of robots to events: Pairwise exchanges and sensor assisted robot
  coordination,'' in \emph{2013 IEEE 10th International Conference on Mobile
  Ad-Hoc and Sensor Systems}, Oct 2013, pp. 249--253.

\bibitem{OlivaSetola2013}
G.~Oliva and R.~Setola, ``Distributed k-means algorithm,'' \emph{CoRR}, vol.
  abs/1312.4176, 2013.

\bibitem{Chung1996}
F.~R.~K. Chung, \emph{Spectral Graph Theory (CBMS Regional Conference Series in
  Mathematics, No. 92)}.\hskip 1em plus 0.5em minus 0.4em\relax American
  Mathematical Society, 1997.

\end{thebibliography}

\end{document}